\documentclass[twocolumn]{aastex701}

\usepackage{enumitem}
\usepackage{amsmath}
\usepackage{silence}

\def\arcsec{\hbox{$^{\prime\prime}$}}
\newcommand{\hst}{\protect\hbox{$H\!ST$} }

\hypersetup{hypertexnames=false}
\usepackage{soul}

\def\II{\,\textsc{ii}}
 
\submitjournal{ApJ}

\shorttitle{Late-time Radio Emission from SN~IIn/II-L}
\shortauthors{Kilpatrick et al.}

\begin{document}

\title{Probing the Mass-loss Histories of Type IIn and II-L Supernovae with Late-time Radio Observations}

\newcommand{\AffCIERA}{\affiliation{Center for Interdisciplinary Exploration and Research in Astrophysics (CIERA), Northwestern University, Evanston, IL 60208, USA}}
\newcommand{\AffNU}{\affiliation{Department of Physics and Astronomy, Northwestern University, 2145 Sheridan Road, Evanston, IL 60208, USA}}
\newcommand{\AffGEM}{\affiliation{Gemini Observatory, 670~North~A'ohoku~Place, Hilo, HI~96720-2700, USA}}
\newcommand{\AffSTScI}{\affiliation{Space Telescope Science Institute, 3700 San Martin Drive, Baltimore, MD 21218, USA}}
\newcommand{\AffUA}{\affiliation{Steward Observatory, University of Arizona, 933 North Cherry Avenue, Tucson, AZ 85721-0065, USA}}

\author[0000-0002-5740-7747]{Charles D.~Kilpatrick} \AffCIERA \email{ckilpatrick@northwestern.edu}

\author[0000-0003-4587-2366]{Lindsay DeMarchi} \AffCIERA\AffNU \email{lindsaydemarchi@gmail.com}

\author[0000-0002-7374-935X]{Wen-fai Fong} \AffCIERA\AffNU \email{wfong@northwestern.edu}

\author[0000-0003-0123-0062]{Jennifer E.~Andrews} \AffGEM \email{jennifer.andrews@noirlab.edu}

\author[0000-0003-2238-1572]{Ori D.~Fox} \AffSTScI \email{ofox@stsci.edu}

\author[0000-0001-5510-2424]{Nathan Smith} \AffUA \email{nathansmith@arizona.edu}

\begin{abstract}

We present VLA observations of 16 Type IIn and Type II-L supernovae (SNe~IIn and SNe~II-L) at $\approx$1000--7000~days after explosion, probing circumstellar matter (CSM) at distances $>10^{16}$~cm from the progenitor corresponding to mass-loss over hundreds to thousands of years before core collapse.  We detect radio emission from four SNe (1998S, 2005ip, 2008fq, and PTF11iqb) with the remaining 12 yielding upper limits of $\nu L_{\nu} < 10^{35}$--$10^{36}~\text{erg s}^{-1}$ at $3$--$11$~GHz.  The detected sources span approximately two orders of magnitude in radio luminosity, reflecting a wide range of CSM densities and pre-explosion mass-loss histories.  All detected sources exhibit steep spectral indices ($\alpha \lesssim -0.4$) consistent with optically-thin synchrotron emission, and the spectral evolution supports internal free-free absorption as the dominant absorption mechanism at these late epochs.  We infer progenitor mass-loss rates of $\dot{M}/v_{w} \lesssim 10^{-6}$--$10^{-3}~M_{\odot}~\text{yr}^{-1}/(100~\text{km s}^{-1})$, with the most radio-luminous objects requiring sustained mass-loss over hundreds to thousands of years.  The detection of the intermediate SN~IIn/SN~II-L object PTF11iqb at luminosities between classical SNe~IIn and SNe~II-L supports a continuum between these subtypes in terms of CSM interaction strength.  Our limits further suggest that SNe~IIn and SNe~II-L are not separated by long-term mass-loss rate at the radii probed here, but chiefly by the presence and strength of dense circumstellar material immediately before explosion.  At epochs $>5000$~days, some SNe (e.g., 1979C and 1986J) maintain nearly constant radio luminosity while others decline rapidly, suggesting that the most radio-luminous SNe~IIn arise from progenitors with sustained mass-loss extending $>10^{4}$~yr before explosion.

\end{abstract}


\section{INTRODUCTION}

Type~IIn supernovae (SNe~IIn) are transient sources that exhibit narrow lines of hydrogen in their optical spectra \citep[for a review, see][]{smith+16,chandra+18,fraser+20}.  These narrow lines arise when ejecta from the explosion of a massive star interact with and shock dense circumstellar matter (CSM) in its immediate environment.  As a result, SNe~IIn efficiently convert kinetic energy into radiation and some of these objects are among the most optically-luminous SNe \citep[e.g., SN~2006gy,][]{smith+07}.
The extremely high luminosities associated with SNe~IIn extend to wavelengths beyond the optical.  Some SNe~IIn emit copious thermal X-ray emission from gas heated behind the SN shock \citep[e.g.,][]{chevalier82,chandra+12,chandra+15} and well-studied examples have been shown to have X-ray luminosities at peak of $>10^{40}~\text{erg s}^{-1}$ \citep[e.g., SN~1988Z,][]{fabian+96}.  Even at mid- and far-infrared wavelengths, thermal emission from newly formed dust in SN-shocked CSM can exceed $10^{42}~\text{erg s}^{-1}$ \citep{fox+11}.

It is therefore curious that radio emission from SNe~IIn appears to exhibit a sharp dichotomy between intrinsically faint and bright sources \citep{boisseau+21}.  While examples such as the SNe\,IIn 1988Z, 1986J, and 2005ip are among the most luminous radio SNe observed \citep{weiler+90,van-dyk+93,williams+02,smith+17}, systematic studies of GHz radio emission from other SNe~IIn have so far yielded mostly upper limits \citep{van-dyk+96}.

Observational studies have shown that emission from all radio SNe is nonthermal \citep[with $\alpha < -0.5$, where flux density $f_{\nu}$ and frequency $\nu$ follow $f_{\nu} \propto \nu^{\alpha}$ as in][]{sramek+80,weiler+82} and thus the emission is thought to come from synchrotron-emitting relativistic electrons \citep{murase+19}.  This mechanism requires both particle acceleration and magnetic field enhancement, and \citet{chevalier82} proposed that Rayleigh-Taylor instabilities created by the deceleration of SN ejecta by the surrounding medium can produce both effects.  These enhancements can be even stronger when the surrounding environment is extremely dense as in SNe~IIn.

However, several possible explanations could account for the large population of SNe~IIn with low radio luminosities.  Although dense environments lead to strong shocks, particle acceleration, and magnetic field amplification, they can also lead to enhanced free-free absorption.

For example, some SNe~IIn exhibit bright H$\alpha$ and infrared emission immediately after explosion, implying extremely high mass-loss rates \citep[$10^{-2}$--$10^{-1}~M_{\odot}~\text{yr}^{-1}$,][]{smith+07,smith08tf,fox+10,fox+11}.  If high mass-loss rates occurred episodically or continuously at a lower level for tens to hundreds of years before explosion, they could also lead to enhanced absorption along the line of sight and obscure any radio signal.  This scenario would require that SNe~IIn explode in complex and structured environments, with electron column densities $>10^{21}~\text{cm}^{-2}$ in order to produce optically-thick absorption at large distances from the progenitor star \citep{weiler+90}.

Optical studies of individual SNe~IIn support the same general picture.  SN~2005ip, for instance, required long-lived pre-supernova mass-loss and dense CSM, perhaps in clumps or a shell, at distances $>0.01$~pc from the progenitor star \citep{smith09,smith+17}.  Taken together, these arguments suggest that observational bias could contribute to a lack of luminous radio SNe~IIn, because most of these objects are observed at early phases when radio emission is likely obscured.  At the same time, the most radio-luminous SNe~IIn are not necessarily representative of the overall observational class: objects such as SN~2005ip and SN~1988Z are unusual in sustaining strong late-time shock-CSM interaction, as traced by narrow H$\alpha$ emission and X-rays, for decades after explosion \citep{fabian+96,van-dyk+93,williams+02,smith09,smith+17,fox+20}, whereas similar strong shock indicators from other SNe~IIn fades much more rapidly at comparable epochs \citep[e.g.,][]{taddia+13,van-dyk+96}.

If early-time radio emission is often hidden in this way, late-time follow-up becomes essential for constraining the mass-loss history of SNe~IIn progenitor stars.  The photosphere and much of the early-time X-ray- and radio-emitting volume can lie within optically-thick CSM.  Observations at epochs of $\gtrsim 10^{3}$--$10^{4}$~days \citep[$\approx$3--30~yr, whereas previous analyses focused on radio emission at $<$5~yr;][]{van-dyk+96,Bietenholz21} are often the most direct way to probe the outer, lower-density wind at radii of $\gtrsim 10^{16}$~cm.  By then the ejecta have expanded enough that free-free optical depths along the line of sight have dropped sufficiently.  Thus intrinsic diversity, together with either structured geometry or line-of-sight obscuration of the radio emission, may contribute to the observed distribution in late-time radio emission from SNe.

A complementary explanation may be the range of CSM densities and thus radio luminosities for SNe~IIn as a whole.  In particular, Type II-L SNe (SNe~II-L) exhibit a long period of linear decline in their light curves, consistent with the explosion of a star with a low-mass hydrogen envelope \citep[see, e.g.,][]{gall+15,bose+18}.  As inferred from their progenitor systems \citep[e.g., SNe~2009kr and 2009hd,][]{fraser+10,elias-rosa+10,elias-rosa+11} and optical spectra \citep{henry+87,chugai+95,milisavljevic+08,terreran+16}, SNe~II-L progenitor stars undergo weaker pre-SN mass-loss than SNe~IIn progenitors \citep[although stronger than SN~II-P progenitors;][]{henry+87,gall+15,morozova+17,bose+18}, and there is spectroscopic evidence that a continuum exists between these two types \citep{smith+15}.  For example, the SNe~II-L 1979C and 1980K were luminous in the optical and radio and both exhibited evidence of strong CSM interactions long after explosion \citep{chevalier+94}.  On the other hand, a progenitor with relatively weak, continuous wind-driven mass loss followed by a strong pre-explosion outburst \citep[as in][]{fox+11} can yield a SN~IIn or SN~II-L that lacks strong radio emission for most of its evolution.  The connection between SNe~IIn and SNe~II-L may also help explain the apparent dichotomy among SNe~IIn radio luminosities: a continuum of CSM interaction strengths, and therefore of radio luminosities, could bridge the gap between intrinsically luminous and faint sources.  In support of this picture, SNe~IIn and SNe~II-L with radio detections span a continuum in peak $5$~GHz luminosity and rise time \citep{weiler+01}, plausibly tracing a continuum in CSM density \citep{weiler+02}.

Recent radio campaigns reinforce the value of late-time monitoring for constraining progenitor mass-loss histories.  Studies of interacting core-collapse SNe have shown that radio data at late phases can trace CSM at radii that are inaccessible at early times and can reveal prolonged or structured mass-loss before explosion \citep{Brennan+25,Tartaglia+25,chandra+18,stroh+21}.  Recent interferometric and VLBI studies also demonstrate that CSM interaction is diverse across SN subtypes, including stripped-envelope events and Type Ibn objects, and that radio morphology and spectral evolution provide direct constraints on CSM density structure \citep{Sollerman+24,Dong+25,Kundu+25}.  These developments motivate a homogeneous late-time study of SNe~IIn and SNe~II-L with consistent observing strategy and analysis.

Here we present observations of 16 SNe~IIn and SNe~II-L with $D\lesssim$50~Mpc (listed in Table~\ref{tab:archival}) using the Karl G.~Jansky Very Large Array (VLA), supplemented at lower frequency ($\sim 3$~GHz) by archival imaging from the Karl G.~Jansky Very Large Array Sky Survey (VLASS, Section~\ref{sec:vlass}).  We compare the radio luminosities derived from these observations to values for other SNe~IIn and SNe~II-L in the literature and discuss the distribution of radio luminosities associated with SNe with strong CSM interactions.  We discuss the constraints on CSM profile and SN ejecta that radio observations can provide and compare these inferences from radio SNe detected in our sample to observations at other wavelengths.

\section{OBSERVATIONS}\label{sec:obs}

\subsection{Sample}\label{sec:sample}

We selected eleven SNe~IIn and five SNe~II-L within $\approx50$~Mpc that either had no radio data or had no published radio follow-up beyond $\sim$2~yr after discovery.  We selected against SNe that were discovered $<5$\arcsec\ from the centers of their host galaxies in order to mitigate confusion with compact nuclear radio cores and bright extended disk emission in their hosts.  The cut does not eliminate all risk of contamination at $\mu$Jy levels: even offset SNe can sit on or near star-forming knots and H\II\ complexes, so association arguments in Section~\ref{sec:connection} rely on astrometric coincidence with the optical transient and nonthermal spectra where detections are secure.  SN~2008fq illustrates a centrally located event where host blending remains important despite optical coincidence \citep{taddia+13,bilinski+15}.  The resulting sample (Table~\ref{tab:archival}) is not volume-limited and spans a range of ages, from $2$--$20$~yr at the time of observation.

\begin{deluxetable*}
	{cccccccc}
	\tabletypesize{\footnotesize}
    \tablecaption{Targets Selected for VLA Observations}
\tablehead{
Supernova &Type& $\alpha$ (J2000) & $\delta$ (J2000) & Discovery Date & Redshift\tablenotemark{a} & Distance & References \\      
          &    &             &            &                   &                                                            & (Mpc)    &     }
\startdata
SN~1996bu & IIn     & 11:20:59.30   & $+$53:12:08.4 & 14 Nov 1996       & 0.003856 & 13     & 1     \\
SN~1998S  & IIn     & 11:46:06.18   & $+$07:28:55.5 & 03 Mar 1998       & 0.002987 & 17     & 2,3     \\
SN~2000P  & IIn     & 13:07:10.53   & $-$28:14:02.5 & 08 Mar 2000       & 0.007542 & 33     & 4,5     \\
SN~2000dc & II-L    & 20:20:45.47   & $-$24:07:57.5 & 09 Aug 2001       & 0.010397 & 40     & 6,7     \\
SN~2001do & II-L    & 19:37:22.76   & $+$40:42:22.8 & 14 Aug 2001       & 0.010421 & 48     & 8,9     \\
SN~2005ip & IIn     & 09:32:06.42   & $+$08:26:44.4 & 05 Nov 2005       & 0.007138 & 31     & 10,11,12   \\
SN~2006am & IIn     & 14:27:37.24   & $+$41:15:35.4 & 22 Feb 2006       & 0.008856 & 44     & 13,14 \\
SN~2008fq & IIn     & 20:25:06.19   & $-$24:48:27.6 & 15 Sep 2008       & 0.010614 & 50     & 15,16 \\
SN~2009ip & IIn     & 22:23:08.20   & $-$28:56:52.6 & 18 Aug 2012\tablenotemark{b} & 0.005944 & 20 & 17,18,19 \\
SN~2009hd & II-L    & 11:20:16.90   & $+$12:58:47.1 & 03 Jul 2009       & 0.002425 & 8      & 20,21   \\
SN~2009kr & II-L    & 05:12:03.25   & $-$15:41:53.1 & 07 Nov 2009       & 0.006468 & 25     & 22,23   \\
SN~2010jp & II-L    & 06:16:30.63   & $-$21:24:36.2 & 11 Nov 2010       & 0.009143 & 36     & 24,25   \\
SN~2011A  & IIn     & 13:01:01.19   & $-$14:31:34.8 & 02 Jan 2011       & 0.008916 & 38     & 26   \\
SN~2011ht & IIn     & 10:08:10.58   & $+$51:50:57.1 & 29 Sep 2011       & 0.003646 & 19     & 27,28,29   \\
PTF11iqb  &IIn/II-L & 00:34:04.84   & $-$09:42:17.9 & 24 Jul 2011       & 0.012499 & 50     & 30,31   \\
SN~2015D\tablenotemark{c}  & IIn     & 13:52:24.11   & $+$39:41:28.2 & 10 Jan 2015       & 0.007222 & 50     & 32,33   \\
\enddata
  \tablecomments{Coordinates, discovery dates, and classifications are from published discovery/classification reports.  Distances are from the Extragalactic Distance Database \citep{tully+09}.  References are  (1): \citep{Nakano6505}, (2): \citep{Li6829}, (3): \citep{Filippenko6830}, (4): \citep{Jha7379}, (5): \citep{Jha7381}, (6): \citep{Yu7476}, (7): \citep{Leonard7483}, (8): \citep{Modjaz7682}, (9): \citep{Chornock7699}, (10): \citep{Boles275}, (11): \citep{Modjaz276}, (12): \citep{Smith695}, (13): \citep{Lee412}, (14): \citep{Blondin8680}, (15): \citep{Thrasher1507}, (16): \citep{Quinn1510}, (17): \citep{Mauerhan430}, (18): \citep{Pastorello767}, (19): \citep{Fraser433}, (20): \citep{Monard1867}, (21): \citep{EliasRosa742}, (22): \citep{Nakano2006}, (23): \citep{Li2042}, (24): \citep{Maza2544}, (25): \citep{Challis2548}, (26): \citep{Pignata2623}, (27): \citep{Boles2851}, (28): \citep{Prieto3749}, (29): \citep{Mauerhan431}, (30): \citep{Parrent3510}, (31): \citep{Smith449}, (32): \citep{Jin4051}, (33): \citep{Zhang6939}.}\label{tab:archival}
  \tablenotetext{a}{These values are derived from the NASA/IPAC Extragalactic Database (\url{https://ned.ipac.caltech.edu/}).}
  \tablenotetext{b}{We report the discovery date as the start of the SN~2009ip-12A event as described in \citet{pastorello+13}.}
  \tablenotetext{c}{Also called PSN J13522411+3941286.}
\end{deluxetable*}

\subsection{VLA Observations}\label{sec:vla-obs}

We observed 16 SNe over at least one epoch between 4 February 2016 and 18 December 2017 (Programs 16A-101, 16A-439, 16B-428, and 17B-201; PI Kilpatrick).  In every epoch, we observed in $C$ and $X$ bands in order to provide contemporaneous measurements or limits at two frequencies.  The time on-source was approximately 14--22~minutes per band.  Some SNe with detections were observed over subsequent epochs.

Our scheduling used VLA configurations A, B, and C (Table~\ref{tab:obs}), corresponding to maximum baselines from $\sim 1$--$36$~km \citep{Perley+11}.  At their representative frequencies, the $C$-band coverage spans roughly $4.9$--$6.2$~GHz and $X$ band roughly $9.7$--$11.0$~GHz, bracketing the $\sim 5$--$10$~GHz range commonly used in the radio SN literature (Section~\ref{sec:archive}).  Typical synthesized beams were $\approx 0.3$--$2$\arcsec\ at $\sim 6$~GHz and $\approx 0.15$--$1.2$\arcsec\ at $\sim 10$~GHz, depending on configuration.  Stokes $I$ images near the SN positions have characteristic rms noise $\sim 3$--$8~\mu$Jy~beam$^{-1}$ at $\sim 6$~GHz for the integration times listed above, sufficient to reach the $\mu$Jy upper limits reported in Table~\ref{tab:obs}.  At the reported distances of our targets, our observations generally reach 3$\sigma$ limiting C- and X-band luminosities of $\nu L_{\nu} = 3\times10^{34}$--$6\times10^{36}$~erg~s$^{-1}$.

\begin{figure*}[t]
	\begin{center}
	\setlength{\fboxsep}{-0.25pt}
      \setlength{\fboxrule}{1.0pt}
      \begin{minipage}{\textwidth}
		\fbox{\includegraphics[width=0.485\textwidth]{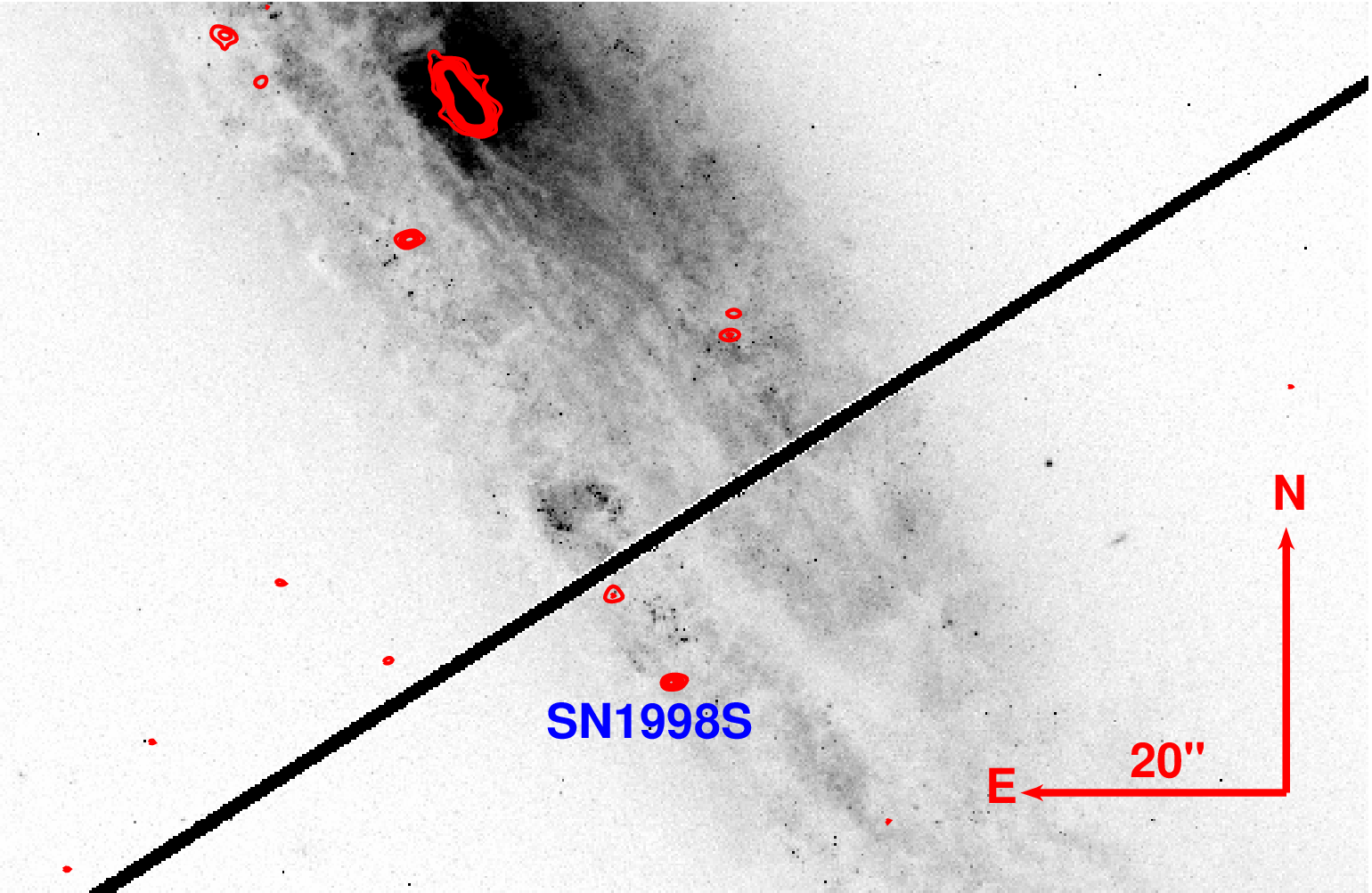}}
		\fbox{\includegraphics[width=0.485\textwidth]{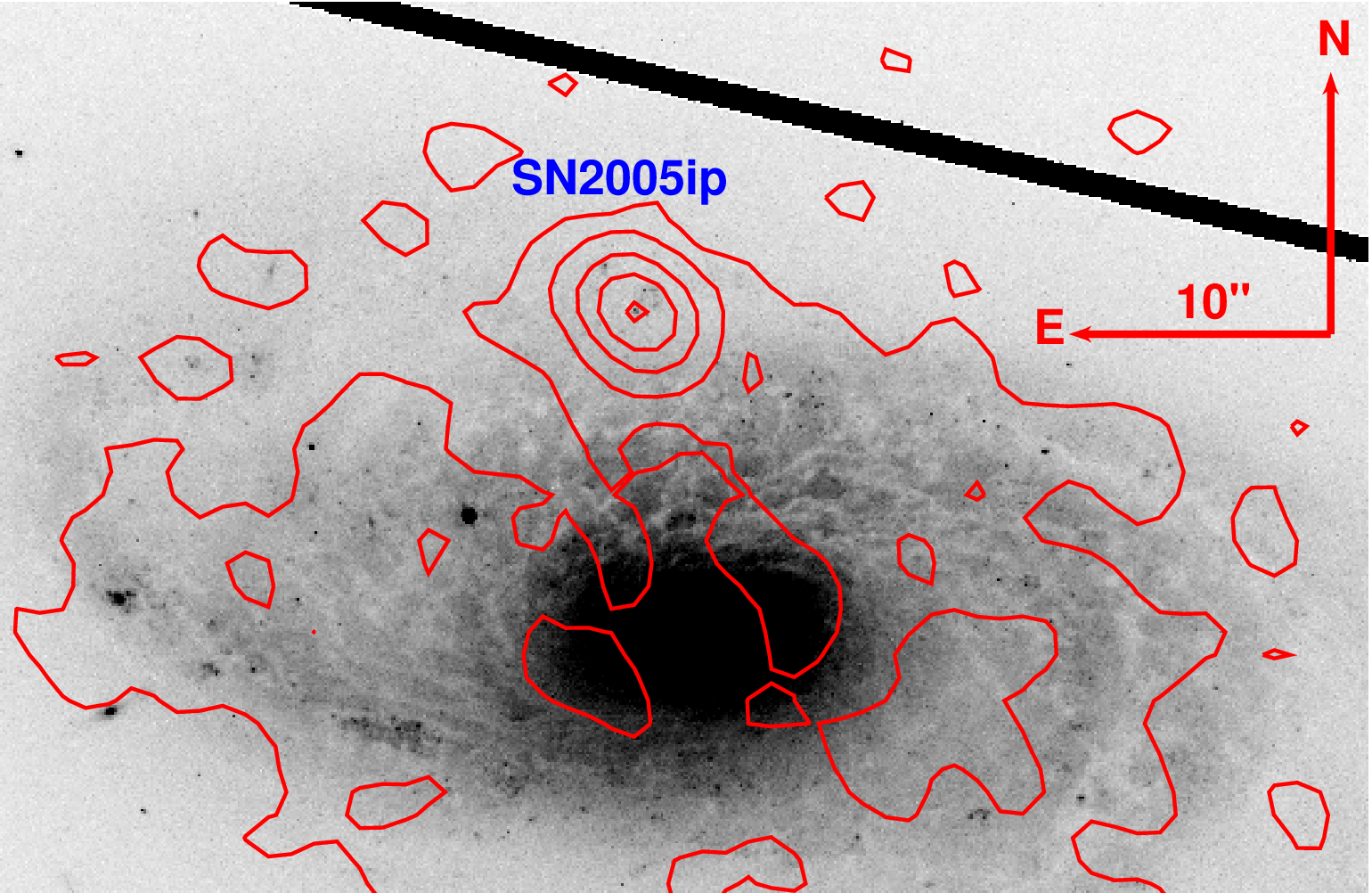}}
	  \end{minipage}
      \begin{minipage}{\textwidth}
		\fbox{\includegraphics[width=0.485\textwidth]{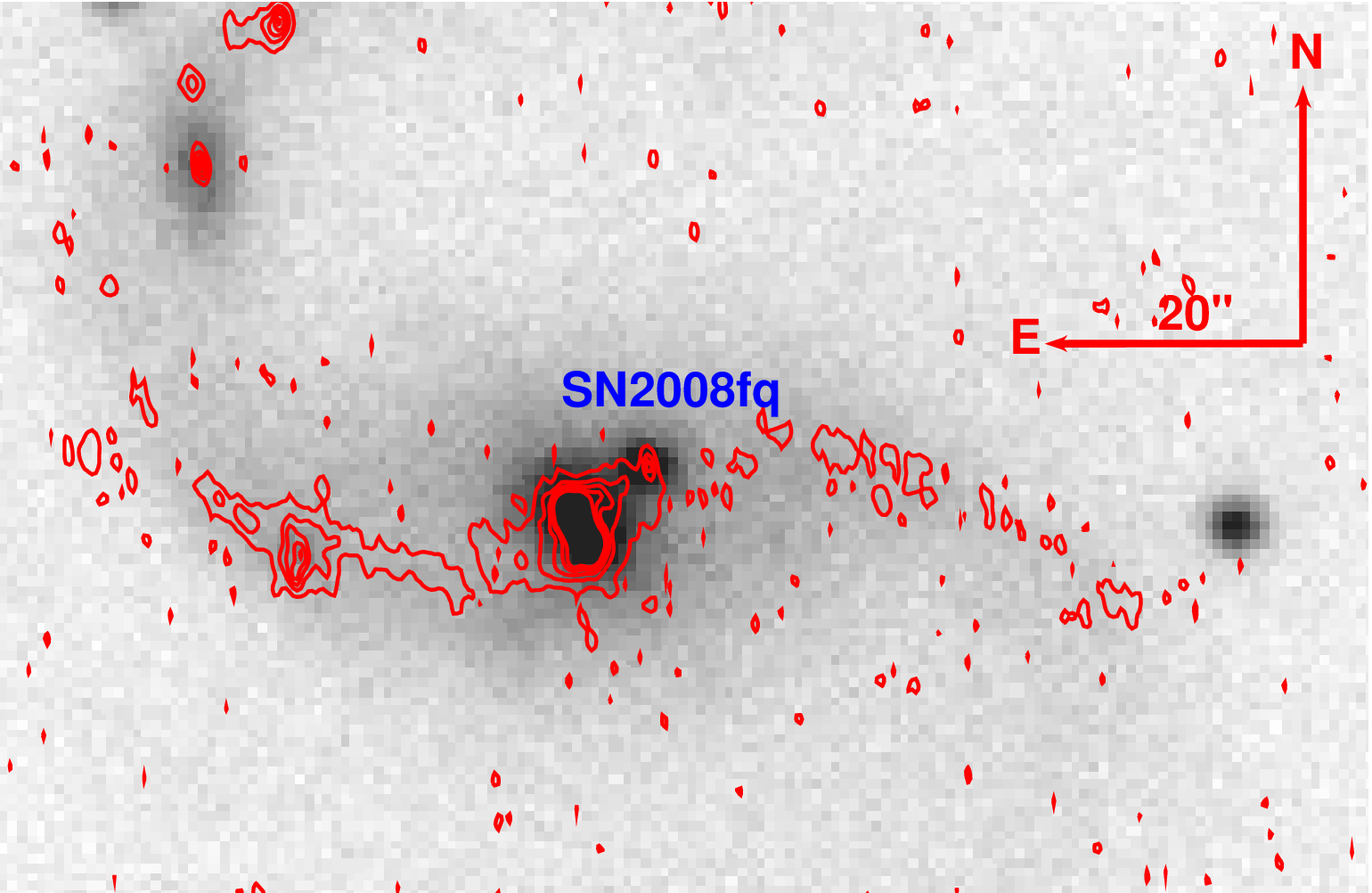}}
		\fbox{\includegraphics[width=0.485\textwidth]{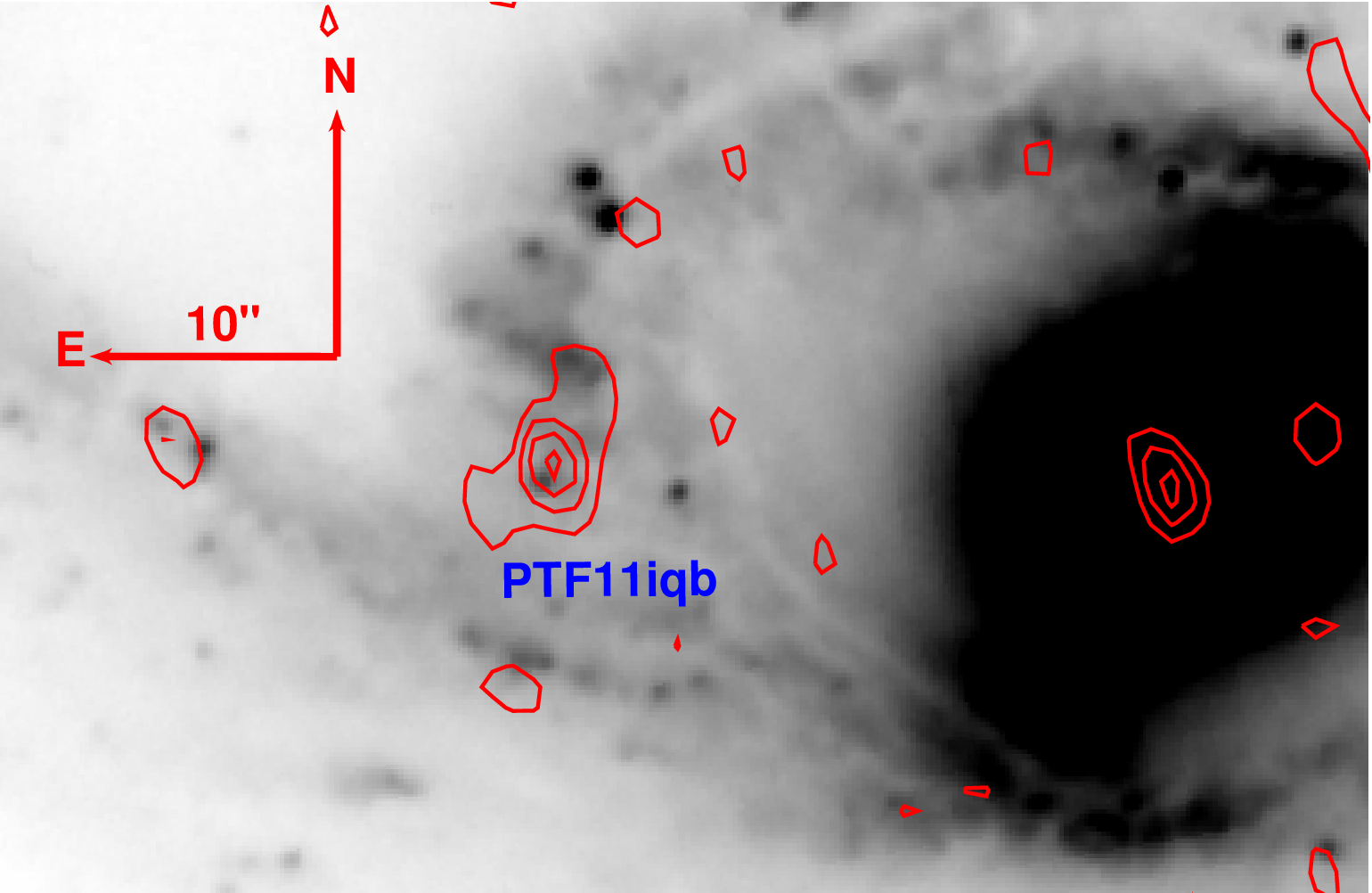}}
		\end{minipage}
      \end{center}
      \setlength{\fboxrule}{0.5pt}
      \vspace{-5pt}
\caption{Four panels showing 6.2~GHz radio contours (red) overlaid on top of optical imaging, as described in Section~\ref{sec:archive} for each SNe with radio sources detected near the approximate position of the optical SN.  The position of each SN is labeled in blue as given in Table~\ref{tab:archival}.  The red contours represent $2.5$, $5$, $10$, and $20$ times the 6.2~GHz rms level.}\label{fig:images}
\end{figure*}

Following standard procedures in the Common Astronomy Software Applications package \citep[CASA,][]{mcmullin+07}, we used the VLA Calibration pipeline to flag and calibrate the data.  In some cases, additional flagging was necessary, and we manually flagged and calibrated these data.  We used CASA/{\tt clean} for deconvolution and imaging with ``Briggs'' weighting of the visibilities and a robustness parameter of $0.5$.  Otherwise, we used default CASA/{\tt clean} parameters.  The deconvolved radio data are shown as contours overlaid on top of optical imaging in Figure~\ref{fig:images}.

For data sets in which we detected a radio source at the approximate position of the SN given in Table~\ref{tab:archival}, we calculated a flux density using CASA/{\tt imfit}.  For high significance in-band detections, we further split the data into sub-bands in order to provide a stronger constraint on the spectral index for that object and epoch.  For non-detections, we calculated a $3\sigma$ upper limit on any radio emission using CASA/{\tt imstat} at the location of the SN and within a region fixed to the size of the Gaussian beam.  In Table~\ref{tab:obs}, we report every observation with the target name, epoch relative to discovery, frequency, and the detected flux density or upper limit.

\subsection{VLASS Archival Imaging and Forced Photometry}\label{sec:vlass}

The Karl G.~Jansky Very Large Array Sky Survey (VLASS) of \citet{lacy+20} is a multi-epoch survey of the sky north of $\delta \approx -40^{\circ}$ at S~band ($\sim$2--4~GHz), with typical synthesized beams of a few arcseconds and characteristic image sensitivities of tens of $\mu$Jy~beam$^{-1}$ in publicly released Quick Look products \citep{gordon+21}.  VLASS is designed in part to characterize radio variability and transients on month-to-year timescales, including stellar explosions and other classes of extragalactic variables \citep{lacy+20,stroh+21}.

To complement the dedicated VLA $C$- and $X$-band observations described in Section~\ref{sec:vla-obs}, we extracted VLASS Quick Look Stokes~$I$ images for each SN position in Table~\ref{tab:archival} from the Canadian Astronomy Data Centre\footnote{\url{https://www.cadc-ccda.hia-iha.nrc-cnrc.gc.ca/en/vlass/}} (CADC).  For each target we queried image cutouts through the CADC DataLink service and retained imaging by VLASS survey epoch (e.g., VLASS1.1--VLASS4.1).  At every epoch we performed forced photometry at the fixed SN sky position from Table~\ref{tab:archival}: we interpolated the Stokes~$I$ brightness (in Jy~beam$^{-1}$) at the tabulated right ascension and declination, estimated the local rms in an annular region outside the central synthesized beam, and measured the brightness or a $3\sigma$ upper limit on the flux density.  All VLASS detections and upper limits are indicated in our complete photometry table (see Appendix~\ref{app:obslog}).

Across the 16 targets, only SN~2005ip shows clear $S/N>3$ detections in the VLASS Quick Look images at $\sim 3$~GHz, with flux densities of order hundreds of $\mu$Jy~beam$^{-1}$ in multiple epochs, qualitatively matching the pattern seen in other late-time radio detections from VLASS \citep{stroh+21}.  This is consistent with the extreme radio luminosity of SN~2005ip relative to the remainder of the sample and with the long-lived, dense circumstellar environment inferred from multiwavelength monitoring \citep{smith+17}.  All other SNe yield $\sim 3\sigma$ upper limits at comparable levels, in line with weak or undetected emission in our higher-frequency VLA imaging.  The VLASS Quick Look products are intended for variability screening rather than flux calibration at the few-percent level \citep{gordon+21}.  Our use of them here is to provide uniform, contemporaneous S-band constraints at the SN positions rather than to replace dedicated VLA follow-up.

\subsection{Other Ancillary Data}\label{sec:archive}

\begin{figure*}[t]
	\begin{center}
		\includegraphics[width=0.94\textwidth]{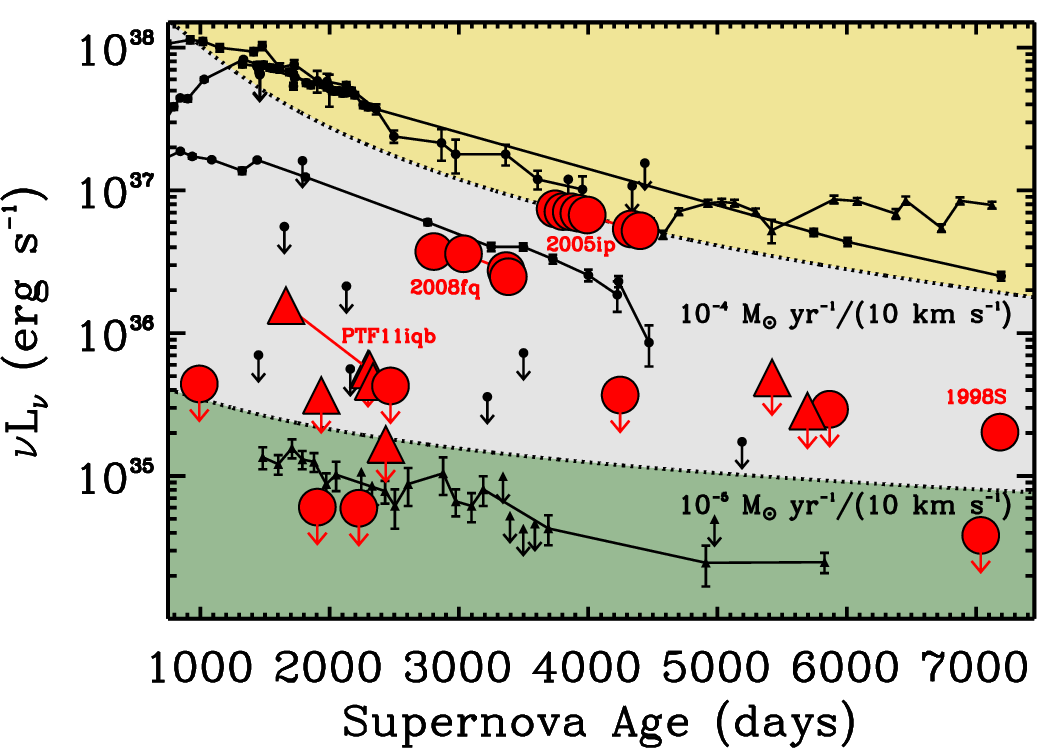}
	\end{center}
\caption{4.86--6.2~GHz radio luminosities from SNe~IIn (indicated with circles) and SNe~II-L (indicated with triangles) as described in Section~\ref{sec:obs}.  Objects in this paper are shown in red while data from the literature are shown in black \citep[from radio light curves and limits in][]{weiler+86,weiler+90,weiler+91,weiler+92,van-dyk+93,eck+96,montes+98,montes+00,bietenholz+02,williams+02,chandra+09,chandra+12}.  Detections for objects in this work are indicated with the name of the object (e.g., for SNe~1998S, 2005ip, 2008fq, and PTF11iqb).  The dashed lines correspond to models describing the light curves of SN~1988Z \citep[from][]{williams+02} scaled such that $\dot{M}/v_{w}=10^{-4}~M_{\odot}~\text{yr}^{-1}/(10~\text{km s}^{-1})$ and SN~1979C \citep{montes+00} scaled to $\dot{M}/v_{w}=10^{-5}~M_{\odot}~\text{yr}^{-1}/(10~\text{km s}^{-1})$.  These light curves roughly divide our sample into three regimes of high luminosity (yellow, SNe~1986J, 1988Z, 2005ip), moderate luminosity (gray, SNe~1995N, 2008fq, PTF11iqb), and low luminosity (green, SN~1980K and various non-detections from this work) radio SNe.}\label{fig:lightcurve}
\end{figure*}

In addition to our own observations, we obtained measurements of SNe~IIn and SNe~II-L at radio wavelengths from several sources in the literature for comparison to our own data. \citet{van-dyk+96} observed $10$ SNe~IIn at $4.86$~GHz, finding upper limits for all ten sources, which we plot along with our own $6.2$~GHz data in Figure~\ref{fig:lightcurve}.  We also obtained $5$~GHz light curves for the SNe~II-L 1979C \citep{weiler+86,weiler+91,montes+00} and 1980K \citep{weiler+92,montes+98} and the SNe~IIn 1986J \citep{weiler+90,bietenholz+02}, 1988Z \citep{van-dyk+93,williams+02}, 1995N \citep{chandra+09}, and 2006jd \citep{chandra+12} as well as a $5$~GHz upper limit for the SN~II-L 1986E \citep{eck+96}, all shown in Figure~\ref{fig:lightcurve}.

The full observation log, including all epochs and in-band frequency splits, is provided in Appendix~\ref{app:obslog} as Table~\ref{tab:obs}.

\section{RESULTS}\label{sec:res}

\subsection{Drawing a Connection Between Supernovae and the Detected Radio Sources}\label{sec:connection}

For four of our targets, we detected radio emission near the approximate location of the optical SN as given in Table~\ref{tab:archival}.  An image is shown for each target in Figure~\ref{fig:images}.  These four detections, SNe~1998S, 2005ip, 2008fq, and PTF11iqb, correspond to a $4/16$ ($25\%$) detection rate among the 16 targets, consistent with the expectation that only a subset of SNe~IIn and SNe~II-L maintain detectable radio emission at late times.  Detection of a radio source near the optical position of a SN may in fact come from the underlying galaxy or a coincident H\II\ region rather than the SN itself.  To confirm the association between the radio sources and the SNe, we assess the likelihood that the emission indeed comes from the SN using two independent methods: (1) correspondence between the optical position of the SN and the position of the peak radio intensity and (2) the spectral index of the radio source.

\subsubsection{Astrometry Between Optical SNe and the Radio Sources}\label{sec:correspondence}

For each detection, we compared the deconvolved radio image to optical imaging of the SN itself.  During our VLA observations, we maintained astrometric accuracy by observing a phase calibrator every 5--6~minutes between observations of the SN (SN~2005ip: J0925+0019, SN~2008fq: J2000-1748, PTF11iqb: J0050-0929, SN~1998S: J1158+4825).  Positional accuracy varies for different VLA configurations and bands, but is typically $<0.15$\arcsec\ in $C$ band.  As a cross-check on our positional accuracy in the deconvolved radio images, we note that the radio contours for each source are coincident with the optical location of the SN and, in each detection, we also detect radio emission from the host galaxy itself (Figure~\ref{fig:images}).

The optical images used for comparison are also shown in Figure~\ref{fig:images}.  These images were obtained with the {\it Hubble Space Telescope} (\hst) WFC3/UVIS in F625W and F814W on 3 Oct 2016 and 11 Jan 2018 for SNe~1998S and 2005ip, respectively (programs SNAP-14668 and SNAP-15166, PI Filippenko).  We reduced these images following procedures described in \citet{kilpatrick+18}.  As we demonstrate in Figure~\ref{fig:images}, the SNe are clearly detected at the positions reported in Table~\ref{tab:archival}.  For SN~2008fq, we used KAIT imaging \citep[described in][]{bilinski+15} of the SN itself from 22 Sep 2008 when the SN was around peak light.  For PTF11iqb, we used $r$-band imaging obtained on 10 Nov 2015 (PI Smith) with the Large Binocular Camera (LBC) on the Large Binocular Telescope (LBT).  This imaging was reduced following standard procedures.  The SN is partly blended with a nearby H\II\ region \citep[described in][]{smith+15}, but we detect it at the position reported in Table~\ref{tab:archival}.

The locations of the radio sources agree with the optical positions of the four SNe shown in Figure~\ref{fig:images}.  For the optical location of SN~2008fq, \citet{taddia+13} and \citet{bilinski+15} emphasize that SN~2008fq is close to the center of the host and there is bright background emission at this location.  Our deconvolved radio image indicates the source we detect is coincident with the optical location of the SN, although the radio emission is partly blended with the host galaxy.  In the other three cases, the radio source and optical SN are clearly coincident.  It is therefore likely that the detected radio sources are associated with the optical SN.

Among our detections, SN~2005ip stands out as the most luminous, with a 5~GHz flux density of $\approx1.1$~mJy at 3744~days post-explosion, declining to $\approx0.8$~mJy by 4401~days.  This corresponds to a luminosity of $\nu L_{\nu} \approx 5 \times 10^{37}~\text{erg s}^{-1}$, making it one of the most luminous radio SNe at this epoch.  SN~2008fq exhibits a more moderate luminosity of $\nu L_{\nu} \approx 3 \times 10^{36}~\text{erg s}^{-1}$, while PTF11iqb shows evidence for significant fading between our first epoch at 1662~days ($f_{\nu} \approx 140~\mu$Jy at 5~GHz) and subsequent epochs at $>2300$~days ($f_{\nu} \approx 25$--$30~\mu$Jy).  SN~1998S, observed at the latest epoch of 7185~days, remains detectable at a flux density of $\approx95~\mu$Jy, demonstrating that radio emission can persist for $\approx 20$~yr after explosion in some SNe~IIn.

\subsubsection{Spectral Indices of the Detected Sources}\label{sec:indices}

\begin{deluxetable}
{ccccc}
\tabletypesize{\small}
\tablecaption{Spectral Indices}
\tablewidth{0pt}
\tablehead{
Supernova & 
Epoch & 
$\alpha$ & 
$f_{{\nu}, 5{\rm GHz}}$   \\       
& 
(days)  
&         
& 
(mJy)
}
\startdata
SN~2005ip & 3743.4 & $-0.44 \pm 0.05$ & 1.07 \\
         & 3804.2 & $-0.48 \pm 0.09$ & 1.03 \\
         & 3868.0 & $-0.54 \pm 0.07$ & 1.02 \\
         & 3927.8 & $-0.55 \pm 0.10$ & 1.00 \\
         & 3991.6 & $-0.54 \pm 0.06$ & 0.97 \\
         & 4332.8 & $-0.48 \pm 0.12$ & 0.81 \\
         & 4400.6 & $-0.51 \pm 0.05$ & 0.79 \\
SN~2008fq & 2805.5 & $-0.89 \pm 0.60$ & 0.24 \\
        & 3035.8 & $-1.21 \pm 0.43$ & 0.24 \\
        & 3364.0 & $-1.05 \pm 0.34$ & 0.18 \\
        & 3381.9 & $-1.24 \pm 0.54$ & 0.17 \\
SN~1998S & 7185.4 & $-0.31 \pm 0.25$ & 0.087 \\
PTF11iqb & 1661.0 & $-1.27 \pm 0.40$ & 0.14 \\
        & 2309.1 & $-1.12 \pm 0.57$ & 0.036 \\
        & 2335.0 & $-0.79 \pm 0.65$ & 0.029 \\
\enddata
\tablecomments{Epoch is since date listed in Table~\ref{tab:archival}.  For high-significance detections, we report the flux density in individual sub-bands.}\label{tab:spectrum}
\end{deluxetable}

Given the flux densities (Table~\ref{tab:spectrum}) in the $C$ and $X$ sub-bands, we find that at our first observing epoch for each source the spectral indices ($\alpha$) of the four detected radio sources are $\alpha=-0.55\pm0.22$ (SN~1998S), $\alpha=-0.40\pm0.03$ (SN~2005ip), $\alpha=-0.89\pm0.59$ (SN~2008fq), $\alpha=-1.27\pm0.38$ (PTF11iqb).  The value of $\alpha$ observed toward all four sources are relatively steep, indicating that the emission arises from a nonthermal source (e.g., synchrotron radiation).

The steep $\alpha$ observed from these sources are also a strong indication that they are not H\II\ regions or otherwise dominated by free-free radio emission.  For a homogeneous H\II\ region at temperature $T_{e}$, the specific intensity at frequency $\nu$ is given by \citep[e.g.,][]{rybicki+79},

\begin{equation}\label{eqn:radtrans}
	I_{\nu} = \frac{2 k T_{e} \nu^{2}}{c^{2}}\left(1 - \exp(-\tau_{\nu})\right).
\end{equation}

\noindent Assuming optical depth $\tau \propto \nu^{-2.1}$ \citep[as in, e.g.,][]{oster+61,mezger+67}, then the $\alpha$ of a free-free emitting cloud will be $-0.1$ in the optically-thin limit and $2$ in the optically-thick limit.  Since we observe sources with $\alpha\lesssim-0.4$, we infer that they cannot come from H\II\ regions or other thermal sources of radio emission.  Rather, the $\alpha$ of these sources are consistent with synchrotron radio emission from SNe.

Furthermore, these $\alpha$ values are consistent with those inferred from observations of other SNe~IIn and SNe~II-L, which become negatively steep at late times and are usually modeled as the evolution from an optically-thick to optically-thin regime \citep{chevalier82,fransson+96,weiler+02,chandra+12,chandra+15}.  In this model, the flux density $f_{\nu}$ at a time after explosion $t$ and with an optical depth $\tau$ can be parameterized in terms of $\alpha, \beta, K_{1}, K_{2}$ as described by \citet{weiler+90,van-dyk+94,weiler+02,chandra+09}:

\begin{eqnarray}
	F_{\nu}(t)&=&K_{1} \left(\frac{\nu}{5~\text{GHz}}\right)^{\alpha} \left(\frac{t}{1000~\text{days}}\right)^{\beta} \exp{(-\tau)} \label{eq:flux} \\
	\tau&=&K_{2} \left(\frac{\nu}{5~\text{GHz}}\right)^{-2.1} \left(\frac{t}{1000~\text{days}}\right)^{\delta}\label{eq:tau}
\end{eqnarray}

\begin{figure}[!t]
	\begin{center}
		\includegraphics[width=\columnwidth]{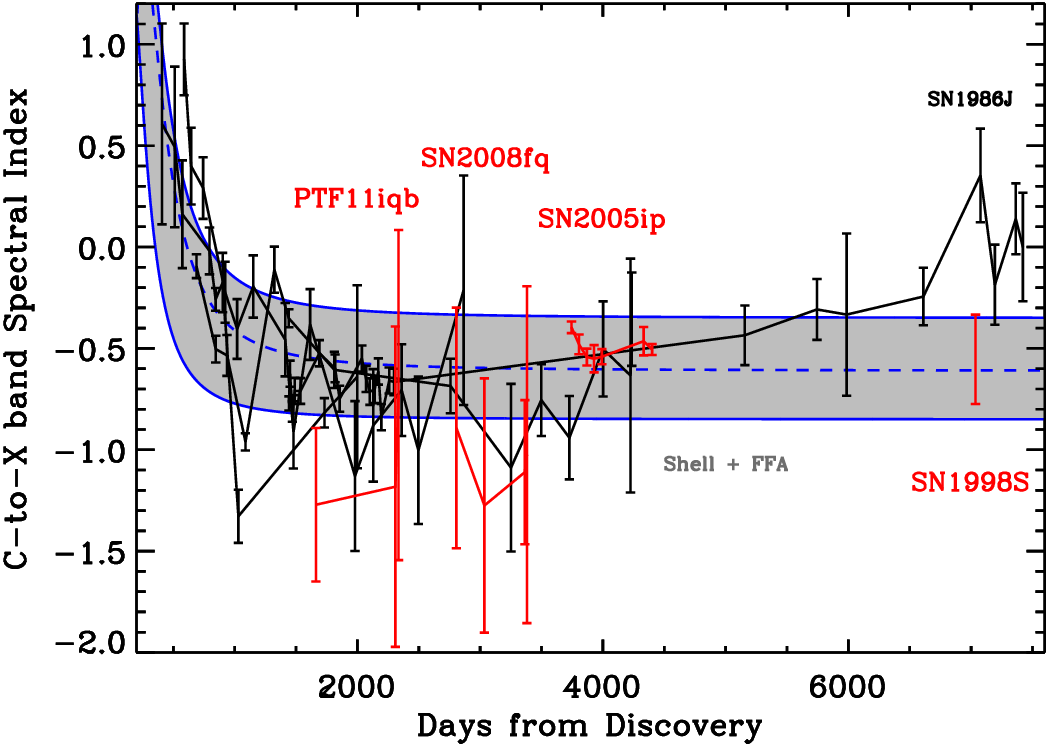}
	\end{center}
\caption{Spectral index from C- to X-bands (roughly $5$--$10$~GHz) as a function of time for SNe~IIn and SNe~II-L.  We label sources described in this paper, including spectral indices derived from the observations described in Section~\ref{sec:indices} (red).  For comparison, we show the theoretical evolution of spectral index for a source dominated by free-free absorption from a shell of dense circumstellar matter, which asymptotically approaches $\alpha=-0.5$, a typical spectral slope for synchrotron emission from optically-thin nonthermal radio SNe and SN remnants \citep{Green19}.}\label{fig:index}
\end{figure}

\noindent Here we assume that radio absorption is dominated by free-free absorption (FFA) from thermal gas mixed with the synchrotron-emitting gas (i.e., internal FFA) as opposed to absorption dominated by an external source of FFA or synchrotron self-absorption \citep[SSA as in][]{chevalier+03}.  This model is preferred for radio light curves of well-studied SNe~IIn, including SNe~1986J \citep{weiler+90}, 1988Z \citep{van-dyk+93,williams+02}, 2006jd \citep{chandra+12}, and 2010jl \citep{chandra+15}.  When the absorption is dominated by internal FFA, the observed spectral index will evolve as:

\begin{equation}
\frac{\partial \log f_{\nu}}{\partial \log \nu} = \alpha - 2.1 \left(\frac{\tau}{\exp{\tau}-1}-1\right)
\end{equation}

\noindent where $\tau$ follows the time-dependent form in Equation~\ref{eq:tau}.  For radio emission from SNe~IIn, the observed spectral index typically varies from positive to negative values as we demonstrate in Figure~\ref{fig:index}.  Here we model all C- to X-band spectral indices assuming $K_{2}=8.11\times10^{5}$, $\delta=-2.45$, and the spectral index asymptotically approaches $\alpha=-0.61$, derived from the median fit to the light curve of SN\,1986J \citep{weiler+90}.  This value is also generally consistent with the $\alpha$ values of young SN remnants presented in \citet{Green19}.

Qualitatively, the radio SNe we have detected exhibit steep spectral indices consistent with a synchrotron-emitting source.  Indeed, PTF11iqb exhibits an extremely steep spectral index at $1662$~days of $-1.27\pm0.40$, which for SNe~IIn and SNe~II-L has only been observed toward SN~2006jd at $1045$--$1742$ days after explosion.  A similar trend was observed in SNe~1986J and 1979C, where a ``deficit'' in high-frequency radio emission (leading to spectral steepening) was interpreted as a variation in thermal absorption from a dense external medium approximately $2000$~days after explosion \citep{weiler+86,lundqvist+88,weiler+90}.  This trend could be a sign that some evolved SNe encounter a pre-supernova wind or shells of CSM at large distances from the progenitor star, consistent with the presence of CSM closer to the progenitor star observed in their optical spectra.

The evolution of spectral indices over time provides additional constraints on the absorption mechanism.  For SN~2005ip, the spectral index remains nearly constant at $\alpha \approx -0.5$ over the $\approx650$~days of our monitoring campaign, suggesting that the source has reached an asymptotic state where absorption effects are minimal.  For SN~2008fq, the spectral index appears to steepen slightly from $\alpha = -0.89$ at 2806~days to $\alpha = -1.24$ at 3382~days, which may indicate that the shock is propagating into a region of decreasing CSM density.  PTF11iqb shows evidence for spectral flattening from $\alpha = -1.27$ at 1662~days to $\alpha = -0.79$ at 2335~days, consistent with the transition from a partially absorbed to an optically-thin synchrotron source.
 
\subsection{Comparison of Radio SN~IIn/II-L Detections and Upper Limits to Values from the Literature}\label{sec:comparison}

In Figure~\ref{fig:lightcurve}, we plot our $6.2$~GHz detections and upper limits from Table~\ref{tab:obs} in terms of luminosity $\nu L_{\nu}=4 \pi D^{2} \nu f_{\nu}$, for a radio SN brightness $f_{\nu}$ at frequency $\nu$ of a source at distance $D$.  VLASS-based luminosities and limits at $\sim 3$~GHz appear in Table~\ref{tab:obs} but are not included in Figure~\ref{fig:lightcurve}, which compares literature and our dedicated $C$-band VLA photometry at $\sim 5$--$6$~GHz.  In addition, we plot luminosities and upper limits of SNe~IIn and SNe~II-L $1500$--$7500$ days after explosion from the literature.  These values are all C-band detections, generally around $4.8$--$5.0$~GHz.

Prior to our work, most SNe~II-L have stringent limits on their radio luminosities of $<10^{36}~\text{erg s}^{-1}$ (with the exception of SN~1979C, discussed below), while most SNe~IIn are at least an order of magnitude more luminous at similar epochs \citep{boisseau+21}, as illustrated in Figure~\ref{fig:lightcurve}.  However, the sample from \citet{van-dyk+96} provides several constraints suggesting some SNe~IIn have intrinsically low radio luminosities.  SNe 1989C, 1989R, and 1988I, observed at 1450, 1650, and 1790 days after optical maximum, respectively, are all less luminous than SNe~IIn detected in the radio at a similar epoch.  At the same time, SNe~IIn detected at these times span at least 2 orders of magnitude in radio luminosity.

The 12 non-detections in our sample provide important constraints on the CSM properties of their progenitors.  Our deepest limits, obtained for SNe~2009ip ($<20.4~\mu$Jy at 6.05~GHz) and SN~2011ht ($<22.2~\mu$Jy at 6.05~GHz), rule out radio luminosities comparable to SN~2005ip or even SN~2008fq at similar epochs.  These stringent limits suggest that the progenitors of these SNe either (1) had significantly lower mass-loss rates in the hundreds of years before explosion, (2) exploded into a more tenuous circumstellar environment, or (3) have CSM geometries that reduce the efficiency of synchrotron emission along our line of sight.  The non-detection of SN~2009ip is particularly noteworthy given the extensive multiwavelength monitoring of this peculiar transient, for which proposed interpretations include a genuine iron core-collapse explosion associated with the 2012 brightening and its broad lines \citep{mauerhan+13ip,pastorello+13}, earlier non-terminal luminous blue variable (LBV) and supernova-impostor activity \citep{smith+10,pastorello+13}, and scenarios in which the major eruptive episodes were not classic iron core-collapse events of a massive star, based in part on limits on synthesized $^{56}$Ni and on the inferred energetics and ejecta properties \citep{fraser+13,margutti+14}.

We now turn to SN~1979C, the SN~II-L exception introduced above in connection with Figure~\ref{fig:lightcurve}, where most SN~II-L are comparatively radio-faint at late times while SN~IIn can extend to much higher luminosities.  SN~1979C is an extremely luminous SN~II-L with persistent optical, X-ray, and radio emission, likely arising from an interaction with dense CSM at large distances from the progenitor star \citep{immler+05}.  This material did not produce narrow H$\alpha$ lines in early time spectra, and only strong, broad H$\alpha$ emission with no P-Cygni absorption was reported \citep{panagia+80,branch+81,barbon+82}.  Narrow H$\alpha$ was discovered in late-time spectra $>10$~yr after discovery \citep{fesen+93}.  These features were interpreted as evidence for dense, unshocked CSM in the environment of the SN~1979C progenitor system.  The CSM inferred for SN~1979C is more similar to that of SNe~IIn at comparable epochs.  In addition, the absence of narrow H$\alpha$ lines at early times \citep[although spectral coverage within the first few years after discovery was variable,][]{panagia+80,branch+81} points to a structured circumstellar environment, with relatively little hydrogen in the immediate environment of its progenitor star.

\subsection{Inferred Mass-Loss Rates}

Various studies have attempted to model the CSM wind density and thus the mass-loss rates for radio SNe by estimating the optical depth to free-free absorption through the CSM and the total luminosity generated by the shock \citep[e.g.,][]{chevalier+82,weiler+86,weiler+91,chevalier+06,Ho+18}.  In practice, this method uses detailed knowledge of the SN radio light curve in order to estimate the optical depth due to FFA or SSA and infer the total column of absorbing gas.  Given the limited information we have from the late-time radio light curves of SNe in our sample, we must estimate the implied mass-loss rate ($\dot{M}$) by making assumptions about various physical parameters intrinsic to SN ejecta and the surrounding CSM.

For the SNe in our sample, the slowly expanding ejecta ($v \leq 0.03c$) are sampled by optical observations.  The SN shock wave propagating into the circumstellar medium carries the fastest-moving ejecta ($v \geq 0.1c$) and emits at radio wavelengths as it encounters material in the surrounding medium \citep{Soderberg+06}.  By monitoring the temporal evolution of the radio spectral peak frequency and peak flux density, one may use radio emission from the interaction of the blast wave and CSM to constrain physical properties of the system, such as $\dot{M}$, the progenitor mass-loss rate \citep{Chevalier98,chevalier+06}.

The physical parameters we derive are summarized in Table~\ref{tab:physical}, assuming microphysical parameters $\epsilon_e = 0.1$ and $\epsilon_B = 0.01$.  These values are motivated by observational constraints from young supernova remnants, which typically find $\epsilon_e > \epsilon_B$.  Studies of cosmic-ray acceleration at SN shocks consistently find that magnetic field amplification is less efficient than electron acceleration, with $10^{-4} \lesssim \epsilon_e \lesssim 0.05$ and $10^{-3} \lesssim \epsilon_B \lesssim 0.1$, yielding ratios of $\epsilon_e/\epsilon_B \sim 10$--$30$ \citep{chevalier+06,Reynolds+21}.  These observational constraints arise from detailed X-ray and radio modeling of well-studied SNRs including Tycho, Kepler, and SN~1006 \citep{Reynolds+21}.  We acknowledge that these are generally lower density than the material around SN\,II-L and IIn.  Assuming equipartition ($\epsilon_e = \epsilon_B = 1/3$) instead would systematically decrease the inferred mass-loss rates by approximately an order of magnitude and increase the inferred shock radii by $\sim$15\%, but would not qualitatively change our conclusions.

We follow the formalism of \citet{DeMarchi22} to model the radio SED and its physical parameters.  Because our $C$- and $X$-band measurements do not span the radio spectral peak at these late epochs, we cannot pin down where the synchrotron spectrum turns over from optically-thick to optically-thin.  For targets with only weak detections or limits we therefore use a synchrotron self-absorption--dominated model as a limiting case for the derived physical bounds.  For example, the telltale signature of synchrotron emission would be $f_{\nu} \propto \nu^{5/2}$ at frequencies $\nu < \nu_{\rm{pk}}$.  Without a direct observation of the optically thick portion of the spectrum, we assume the system is emitting via synchrotron emission.  When the peak is not observed, we treat the lowest observed band as an effective upper bound on $\nu_{\rm{pk}}$.

Information regarding the physical parameters of the system such as magnetic field $B$, energy density $U$ corresponding to $\rho_{\rm{CSM}}v_{FS}^2$ (where $v_{FS}$ is the velocity of the forward shock) and shock wave radius $R$ are derived from the SED peak location and the slope of the optically-thin spectrum.  SNe with multiple epochs of observation were fit jointly as the slopes of the SED are not assumed to vary greatly in time.

Some SN data sets were sparse and only included flux upper limits provided by the instrument.  We assume the emission that would be present from the supernova at such late times would be only caused by optically-thin material, that is, the peak is sufficiently evolved to low frequencies and we see no evidence for CSM interaction at other wavelengths.  An additional assumption that must be made for modeling upper limits with minimal free parameters is to assume all the emission is caused by synchrotron self-absorption such that the equations of \citet{chevalier+82} can be applied.  We note that late-time radio observations of SNe~IIn typically favor internal FFA over SSA as the dominant absorption mechanism \citep{weiler+90,van-dyk+93,williams+02,chandra+12}, which may lead to a slightly lower mass-loss rate than in the FFA case \citep[e.g.,][]{Wu25}.  However, the SSA formalism provides a useful limiting case for constraining physical parameters from our upper limits.

The interaction between the SN ejecta and CSM can be modeled with a double power-law density profile, $\rho_{\rm{CSM}} \propto R^{-n}$ and $\rho_{\rm{wind}} \propto R^{-s}$ \citep{chevalier82}.  The self-similar solutions of \citet{chevalier+82} are characterized by $q \equiv \frac{n-3}{n-s}$, which describes the evolution of the blast wave radius in time $R \propto t^q$.  We adopt $s=2$ as is typical for a wind profile \citep{chevalier+82}, and $n=12$, yielding a value of $q=0.9$ for a convective RSG progenitor density profile \citep{Chevalier01}.  Though we do not know the progenitor unambiguously, this choice of $n$ (and by extension, $q$) encompasses more compact progenitor choices such as WR stars and the progenitor of SN\,1987A \citep[][thus providing a true upper limit]{Chevalier01}.  The largest source of uncertainty on our limits comes from the choice of total energy partition $\epsilon$ (see Appendix C in \citealt{DeMarchi22}).

Knowledge of $B, U, R,$ and $q$ allows us to calculate the density of the CSM $\rho_{\rm{CSM}}$, the number density of electrons $n_e$, and mass loss as a function of the wind velocity $v_{\rm{wind}}$ and wind density parameter $\dot{M}/v_{\rm{wind}}$ \citep[the latter also referred to as $w$;][]{Chevalier01}. We present these values in Table~\ref{tab:physical} assuming microphysical parameters $\epsilon_e=0.1$ and $\epsilon_B = 0.01$, consistent with observational constraints from young supernova remnants that consistently find $\epsilon_e > \epsilon_B$ \citep{chevalier+06,Reynolds+21}.

In a few instances, an estimate of the slope constrained by two upper limits was too shallow ($\alpha > -0.5$), too steep ($\alpha < -2.5$), or ascending (non-physical, but a side effect of a worse constraint at higher frequencies).  In these situations, demarcated by a dagger ($\dagger$) in Table~\ref{tab:physical}, we calculate physical parameters for a range of slopes seen in late-time radio SNe~IIn, $-0.7 < \alpha < -0.5$ \citep{chandra+12, chandra+09, Bietenholz+10, williams+02} and report parameter limits that are the most inclusive within the adapted slope range.

In the specific case of SN~2005ip, the best-fitting slope was $\alpha = -0.50 \pm 0.02$, which approaches the asymptotic value expected for optically-thin synchrotron emission.  In this regime, much like the SNe whose upper limits yield a slope $\alpha > -0.5$, calculations of certain physical parameters become less constraining due to the weak dependence on spectral index in the optically-thin limit.

\begin{deluxetable*}{lccccccc}
\tablecaption{Physical parameters derived from SED modeling using the formalism of \citet{DeMarchi22}, assuming microphysical parameters $\epsilon_e=0.1$ and $\epsilon_B = 0.01$.  Objects marked with $\dagger$ have spectral indices constrained to the range $-0.7 < \alpha < -0.5$ (see text).}
\label{tab:physical}
\tablewidth{0pt}
\tablehead{
\colhead{Name} &
\colhead{Epoch} &
\colhead{$R$} &
\colhead{$n_e$} &
\colhead{$\dot{M}/v_{\rm{w}}$} &
\colhead{$v_{\rm{FS}}$} &
\colhead{$\rho_{\rm{CSM}}$} &
\colhead{$\alpha$} \\
\colhead{} &
\colhead{(days)} &
\colhead{($10^{15}$~cm)} &
\colhead{(cm$^{-3}$)} &
\colhead{($M_{\odot}$~yr$^{-1}$)/(100 km s$^{-1}$)} &
\colhead{($10^{7}$ cm s$^{-1}$)} &
\colhead{(g cm$^{-3}$)} &
\colhead{}
}
\startdata
SN~1996bu$^\dagger$ & 7036 & $\geq$1.08 & $\leq 1.1\times 10^{12}$ & $\leq 5.4\times 10^{-6}$ & $\geq$0.18 & $\leq 1.8\times 10^{-12}$ & $-0.7$ to $-0.5$ \\
SN~1998S & 7185 & $\geq$0.90 & $\leq 5.0\times 10^{11}$ & $\leq 1.4\times 10^{-5}$ & $\geq$0.15 & $\leq 8.3\times 10^{-13}$ & $-0.53$ \\
SN~2000P$^\dagger$ & 5867 & $\geq$1.10 & $\leq 7.3\times 10^{11}$ & $\leq 3.7\times 10^{-6}$ & $\geq$0.22 & $\leq 1.2\times 10^{-12}$ & $-0.7$ to $-0.5$ \\
SN~2000dc$^\dagger$ & 5696 & $\geq$1.09 & $\leq 7.1\times 10^{11}$ & $\leq 3.5\times 10^{-6}$ & $\geq$0.22 & $\leq 1.2\times 10^{-12}$ & $-0.7$ to $-0.5$ \\
SN~2001do$^\dagger$ & 5422 & $\geq$1.40 & $\leq 3.4\times 10^{11}$ & $\leq 2.8\times 10^{-6}$ & $\geq$0.30 & $\leq 5.7\times 10^{-13}$ & $-0.7$ to $-0.5$ \\
SN~2005ip & 3744 & $\geq$6.18 & $\leq 9.2\times 10^{8}$ & $\leq 1.2\times 10^{-5}$ & $\geq$1.91 & $\leq 1.5\times 10^{-15}$ & $-0.52$ \\
SN~2005ip & 4401 & $\geq$5.31 & $\leq 1.9\times 10^{9}$ & $\leq 1.8\times 10^{-5}$ & $\geq$1.40 & $\leq 3.1\times 10^{-15}$ & $-0.52$ \\
SN~2006am$^\dagger$ & 4247 & $\geq$1.26 & $\leq 2.6\times 10^{11}$ & $\leq 1.8\times 10^{-6}$ & $\geq$0.34 & $\leq 4.4\times 10^{-13}$ & $-0.7$ to $-0.5$ \\
SN~2008fq & 2806 & $\geq$4.82 & $\leq 1.8\times 10^{9}$ & $\leq 1.4\times 10^{-5}$ & $\geq$1.99 & $\leq 3.0\times 10^{-15}$ & $-1.04\pm0.06$ \\
SN~2008fq & 3381 & $\geq$4.05 & $\leq 3.8\times 10^{9}$ & $\leq 2.1\times 10^{-5}$ & $\geq$1.39 & $\leq 6.4\times 10^{-15}$ & $-1.04\pm0.06$ \\
SN~2009ip$^\dagger$ & 1905 & $\geq$0.54 & $\leq 4.5\times 10^{11}$ & $\leq 5.4\times 10^{-5}$ & $\geq$0.33 & $\leq 7.4\times 10^{-13}$ & $-0.7$ to $-0.5$ \\
SN~2009hd$^\dagger$ & 2434 & $\geq$0.77 & $\leq 3.1\times 10^{11}$ & $\leq 7.6\times 10^{-5}$ & $\geq$0.37 & $\leq 5.2\times 10^{-13}$ & $-0.7$ to $-0.5$ \\
SN~2009kr$^\dagger$ & 2297 & $\geq$1.50 & $\leq 5.2\times 10^{10}$ & $\leq 4.9\times 10^{-5}$ & $\geq$0.75 & $\leq 8.8\times 10^{-14}$ & $-0.7$ to $-0.5$ \\
SN~2010jp$^\dagger$ & 1935 & $\geq$1.23 & $\leq 6.1\times 10^{10}$ & $\leq 3.8\times 10^{-5}$ & $\geq$0.73 & $\leq 1.0\times 10^{-13}$ & $-0.7$ to $-0.5$ \\
SN~2011A$^\dagger$ & 2471 & $\geq$1.35 & $\leq 7.5\times 10^{10}$ & $\leq 5.7\times 10^{-5}$ & $\geq$0.63 & $\leq 1.3\times 10^{-13}$ & $-0.7$ to $-0.5$ \\
SN~2011ht$^\dagger$ & 2226 & $\geq$0.53 & $\leq 6.2\times 10^{11}$ & $\leq 7.4\times 10^{-5}$ & $\geq$0.28 & $\leq 1.0\times 10^{-12}$ & $-0.7$ to $-0.5$ \\
PTF11iqb & 1662 & $\geq$5.33 & $\leq 4.9\times 10^{8}$ & $\leq 4.6\times 10^{-5}$ & $\geq$3.71 & $\leq 8.2\times 10^{-16}$ & $-1.18\pm0.11$ \\
PTF11iqb & 2335 & $\geq$2.10 & $\leq 1.3\times 10^{10}$ & $\leq 1.8\times 10^{-5}$ & $\geq$1.04 & $\leq 2.1\times 10^{-14}$ & $-1.18\pm0.11$ \\
SN~2015D$^\dagger$ & 992 & $\geq$1.37 & $\leq 1.2\times 10^{10}$ & $\leq 9.1\times 10^{-4}$ & $\geq$1.60 & $\leq 2.0\times 10^{-14}$ & $-0.7$ to $-0.5$ \\
\enddata
\tablecomments{Epoch is days since discovery date listed in Table~\ref{tab:archival}.  For SNe with multiple epochs, we show representative first and last epochs.  All values are upper or lower limits as indicated.}
\end{deluxetable*}

\section{DISCUSSION}\label{sec:disc}

The results presented in the previous section reveal that late-time radio emission from SNe~IIn and SNe~II-L is more complex than a simple luminosity dichotomy would suggest \citep{boisseau+21,chandra+18}.  Our detections and upper limits, combined with data from the literature, point toward a continuum of radio luminosities that reflects the diversity of CSM environments surrounding SN progenitors.  In this section, we discuss the implications of these findings for our understanding of massive star mass-loss in the final stages of stellar evolution.

\subsection{The Continuum of Radio Luminosities in SNe~IIn and SNe~II-L}

Our late-time radio observations reveal that SNe~IIn and SNe~II-L span a wide range of radio luminosities at epochs $>1000$~days after explosion.  This continuum of luminosities, ranging from $\nu L_{\nu} \approx 10^{35}$--$10^{38}~\text{erg s}^{-1}$, reflects the diversity of CSM densities surrounding their progenitors.  The most radio-luminous objects in our sample (SNe~2005ip and 2008fq) have inferred CSM densities comparable to those of the SNe\,IIn 1988Z and 1986J, while the non-detections constrain CSM densities to values more typical of SNe~II-L such as SN~1980K.

This diversity may arise from variations in the pre-explosion mass-loss histories of the progenitor stars.  Each of our detected SNe provides unique insights into these mass-loss histories:

\textit{SN~2005ip} is a SN\,IIn and remains one of the most luminous radio SNe at $>4000$~days post-explosion, with $\nu L_{\nu} \approx 5 \times 10^{37}~\text{erg s}^{-1}$.  The sustained radio emission requires ongoing shock interaction with dense CSM, implying mass-loss extending over $\sim10^{3}$~yr before core collapse \citep{smith09,smith+17}.  \citet{fox+20} presented panchromatic follow-up through $\sim$5000~days post-explosion and argued that the shock--CSM interaction was entering a slow decline even as integrated energetics and CSM properties continued to imply extreme, long-lived progenitor mass loss.  Our measurements show a decline of $\approx25\%$ in flux density over the 657~days of our monitoring campaign, corresponding to a temporal decline rate of $\beta \approx -0.5$, shallower than the late-time decline of synchrotron luminosity expected for a steady wind with $\rho \propto r^{-2}$ in standard radio-CSM interaction models \citep{chevalier-fransson+06araa,weiler+02}.

\textit{SN~2008fq} exhibits radio emission at a level intermediate between SN~2005ip and our non-detections.  The steep spectral indices we measure ($\alpha \approx -1.0$ to $-1.2$) suggest that free-free absorption remains significant even at these late epochs, possibly indicating a denser or more extended CSM than that of typical SNe~IIn.

\textit{PTF11iqb}, classified as an intermediate SN~IIn/SN~II-L object, shows the most dramatic evolution in our sample, fading by a factor of $\approx5$ between 1662 and 2335~days.  This rapid decline may indicate that the shock is exiting a region of dense CSM and propagating into more tenuous material, consistent with the interpretation that PTF11iqb's progenitor experienced a brief episode of enhanced mass-loss before explosion.

\textit{SN~1998S}, observed at the latest epoch in our sample (7185~days), demonstrates that radio emission can persist for $\approx 20$~yr after explosion.  The relatively flat spectral index ($\alpha = -0.31 \pm 0.25$) suggests minimal absorption at these late times.

Sustained mass loss over $\sim10^{3}$~yr exceeds the duration of late nuclear burning phases (Ne, O, and Si burning) and likely corresponds to episodic eruptions during C or He burning \citep{smith+16}.  Alternatively, binary interaction may drive pre-SN mass loss on these timescales \citep{sa14}. Forward- and reverse-shock modeling of SNe~IIn light curves likewise points to a wide range of CSM densities set by pre-explosion mass loss \citep{taddia+19}.  Recent studies of other interacting SNe support this picture: \citet{Brennan+25} found evidence for ongoing ejecta-CSM interaction in SNe~IIn at radii $>10^{16}$~cm, indicating progenitor mass-loss rates of $\sim10^{-4}$--$10^{-5}~M_{\odot}~\text{yr}^{-1}$ in the final hundreds to thousands of years before explosion.  In contrast, the non-detections in our sample may arise from progenitors that experienced only brief periods of enhanced mass-loss immediately before explosion, producing dense CSM only in the immediate environment of the star.

\subsection{Spectral Indices and Absorption Mechanisms}

The spectral indices we measure for the detected radio sources provide important constraints on the physical conditions in the emission region.  All four detected sources exhibit negative spectral indices ($\alpha \lesssim -0.4$), consistent with optically-thin synchrotron emission from relativistic electrons accelerated at the SN shock front.

The evolution of spectral indices in our sample is well-described by a model where internal FFA dominates over SSA at these late epochs.  As demonstrated in Figure~\ref{fig:index}, the observed spectral indices are consistent with an asymptotic approach to $\alpha \approx -0.5$, characteristic of synchrotron emission with minimal absorption.  This finding is consistent with the interpretation that radio emission at $>1500$~days post-explosion samples CSM at distances $>10^{16}$~cm from the progenitor, where electron column densities are insufficient to produce significant external FFA \citep{chandra+18,chandra+15}.

The particularly steep spectral index of PTF11iqb ($\alpha = -1.27 \pm 0.40$) at 1662~days is notable.  Such steep indices have been observed in other SNe~IIn, including SN~2006jd \citep{chandra+12}, and may indicate interaction with a dense shell or wind at large radii from the progenitor.  This interpretation is consistent with PTF11iqb's classification as an intermediate SN~IIn/SN~II-L object, where CSM interaction may be more variable than in classical SNe~IIn \citep{smith+15,Tartaglia+21}.  Similar spectral evolution has been observed in other stripped-envelope SNe with CSM interaction, such as SN~2019oys, where free-free absorption signatures indicated CSM densities of $\sim10^{5}~\text{cm}^{-3}$ and mass-loss rates of $\sim10^{-3}~M_{\odot}~\text{yr}^{-1}$ \citep{Sollerman+24}.

\subsection{Identifying the Progenitors of SN\,IIn and II-L From Progenitor Mass-loss Histories}

The mass-loss range $\dot{M}/v_{\rm w} \lesssim 10^{-6}$--$10^{-3}\,M_{\odot}\,{\rm yr}^{-1}/(100\,{\rm km\,s}^{-1})$ from our analysis matches, to order of magnitude, independent inferences from optical line and light-curve modeling of SNe~IIn and interacting SNe~II-L \citep{smith+17,taddia+19,Tartaglia+21}.  The same range is consistent with X-ray luminosities and absorption in CSM shocks, which together with shock speeds yield comparable $\dot{M}/v_w$ for luminous interactors \citep{chandra+15,immler+05}.  Early-time GHz radio light-curve modeling of the brightest radio SNe similarly constrains $\dot{M}/v_w$ from free-free optical depth and shock energetics \citep{weiler+02,Ho+18,chevalier-fransson+06araa}.  In that multiwavelength context, our late-time radio limits imply that SNe~IIn and SNe~II-L are not cleanly separated by long-term progenitor mass-loss rate. In that multiwavelength context, our late-time radio limits imply that SNe~IIn and SNe~II-L are not cleanly separated by the time-averaged progenitor mass-loss rate inferred on $\sim10^{2}$--$10^{3}$~yr timescales \citep{smith+17,taddia+19,Tartaglia+21,chandra+15,immler+05,weiler+02,Ho+18,chevalier-fransson+06araa}.  Both classes span a wide range of $\dot{M}/v_{\rm w}$, from the faint end consistent with our non-detections to the high end inferred for radio- and X-ray-bright outliers such as SN~2005ip and SN~1979C.  What appears to distinguish them more sharply is instead the presence, strength, and radial extent of dense CSM in the immediate pre-explosion environment: SNe~IIn require enough circumstellar material on months-to-years scales to sustain narrow, shock-ionized line emission \citep{schlegel+90,chevalier+94,chandra+18,fraser+20}, often linked to eruptive or super-Eddington mass loss shortly before core collapse \citep{smith+07,smith08tf,fox+10,fox+11,smith+10}.  By contrast, many SNe~II-L can be modeled as ordinary red-supergiant explosions with only modest steady winds, yet still admit dense CSM in some cases \citep{morozova+17,dessart+23,gall+15,chevalier+94}.  It is therefore notable that, once the shock has reached radii $\gtrsim10^{16}$~cm probed at late times, the inferred long-term mass-loss rates of detected and undetected objects in both classes overlap rather than forming two distinct populations.  Hybrid events such as PTF11iqb occupy intermediate $\dot{M}/v_{\rm w}$ and radio luminosity \citep{smith+15,Tartaglia+21}, supporting a continuum in CSM interaction strength rather than a strict division between SNe~IIn and SNe~II-L based on wind history alone.

The radio-detected SNe in our sample have inferred mass-loss rates that place their progenitors among stars with some of the highest known mass-loss rates.  Our radio limits are consistent with multiple episodes of enhanced mass-loss in the hundreds to thousands of years before explosion, rather than a single steady wind, because the CSM radii probed at late times typically require material ejected over comparable intervals.

Independent multiwavelength studies of nearby SNe~IIn reach similar orders of magnitude and emphasize mechanisms that can repeat over a stellar lifetime.  For SN~2020ywx, \citet{Tartaglia+25} inferred progenitor mass-loss rates of $10^{-2}$--$10^{-3}~M_{\odot}~\text{yr}^{-1}$ sustained for $>100$~yr pre-explosion and highlighted binary interaction as one driver of such histories.  Demographic studies of massive-star binaries further show that Roche-lobe overflow and tidal interaction are common before core collapse \citep{Sana+12}, qualitatively matching the repeated mass-ejection episodes needed to explain luminous late-time radio SNe.

Several classes of evolved massive stars are known to sustain the kinds of long-term, high $\dot{M}$ inferred above, so the progenitors of radio-luminous SNe~IIn and interaction-dominated SNe~II-L need not be unique.  LBVs combine quiescent winds at $\dot{M}\sim10^{-5}$--$10^{-4}~M_{\odot}~\text{yr}^{-1}$ with giant eruptions that can briefly approach $\dot{M}\sim0.1$--$1~M_{\odot}~\text{yr}^{-1}$, building the densest immediate CSM while stars cross the upper instability strip \citep{humphreys+davidson+94,smith+16}.  Direct connections between precursor variability, giant outbursts, and LBV-like transients further motivate this channel \citep{smith+10,kilpatrick+18}.

Yellow and cool hypergiants likewise exhibit strongly time-variable winds and episodic ejection, offering a path to extended CSM over hundreds to thousands of years \citep{humphreys+davidson+94}.  Empirical scalings and atmospheric models for dusty red supergiant winds yield time-averaged $\dot{M}$ comparable to the faint end of our range, with the most luminous objects reaching higher effective mass-loss when clumping, pulsation, and dust are included \citep{vanLoon+05,josselin+07,chevalier+06}.  On the other hand, hot and compact systems such as Wolf-Rayet stars maintain fast, dense radiatively driven winds \citep{crowther+07}.  

Together with evidence that some transitional SNe~IIn/SNe~II-L favor cool-supergiant progenitors \citep{smith+15,taddia+19}, these channels suggest our radio-detected objects plausibly originate from a mix of LBV-like eruptive histories, extreme hypergiants or RSGs, and interacting binaries, rather than a single stellar type.

\subsection{Connection Between SNe~IIn and SNe~II-L}

Our observations support the hypothesis that SNe~IIn and SNe~II-L form a continuum in terms of CSM interaction strength \citep[e.g.,][]{chevalier82,chevalier+94,weiler+02,morozova+17,dessart+23}.  The detection of radio emission from PTF11iqb, classified as an intermediate SN~IIn/SN~II-L object \citep{smith+15,Tartaglia+21}, at luminosities intermediate between classical SNe~IIn (e.g., SN~2005ip) and upper limits for SNe~II-L strengthens this interpretation.

Similarly, SN~1979C, historically classified as a SN~II-L, remains one of the most radio-luminous SNe known despite lacking narrow H$\alpha$ emission at early times \citep{weiler+91}.  The late appearance of narrow emission lines in SN~1979C \citep{fesen+93} suggests that the distinction between SNe~IIn and SNe~II-L may depend on the distribution of CSM rather than its total mass.  Progenitors with extended, low-density CSM may produce SNe~II-L spectra while still being capable of sustained radio emission at late times.

The continuum in observed mass-loss histories at $>5000$~days, where some SNe (e.g., SNe~1979C and 1986J) maintain nearly constant radio luminosity while others decline rapidly, may reflect the presence or absence of pre-supernova mass-loss at correspondingly earlier epochs ($>10^{4}$~yr before explosion).  This timescale corresponds to the main-sequence or early post-main-sequence evolution of the progenitor, suggesting that the most radio-luminous SNe~IIn may arise from stars that experienced episodic mass-loss throughout much of their evolution.

\section{CONCLUSIONS}\label{sec:conclusion}

We have presented VLA observations of 16 SNe~IIn and SNe~II-L at epochs ranging from $\approx1000$--$7000$~days after explosion, together with archival VLASS Quick Look photometry at $\sim 3$~GHz for each target (Section~\ref{sec:vlass}).  These late-time observations probe CSM at distances of $>10^{16}$~cm from the progenitor star, corresponding to mass-loss that occurred hundreds to thousands of years before core collapse \citep{chandra+18,stroh+21}.  By combining our new observations with data from the literature, we synthesize a broad view of the radio properties of SNe~IIn and SNe~II-L at late times ($\gtrsim 10^{3}$~days), complementing earlier work focused on individual objects and smaller samples.  Our principal findings are:

\begin{enumerate}[leftmargin=*]
\item We detect radio emission from four SNe: SN~1998S, SN~2005ip, SN~2008fq, and PTF11iqb.  The remaining 12 targets yield upper limits that constrain their radio luminosities to $\nu L_{\nu} < 10^{35}$--$10^{36}~\text{erg s}^{-1}$.

\item The detected sources span approximately two orders of magnitude in radio luminosity at similar epochs, with SN~2005ip being among the most luminous radio SNe known at $>3700$~days post-explosion.  This diversity reflects a wide range of CSM densities at large distances ($>10^{16}$~cm) from the progenitor stars.

\item All detected sources exhibit steep spectral indices ($\alpha \lesssim -0.4$) consistent with optically-thin synchrotron emission.  The evolution of spectral indices in our sample supports internal free-free absorption as the dominant absorption mechanism at these late epochs.

\item We infer episodic progenitor mass-loss rates of $\dot{M}/v_{w} \lesssim 10^{-6}$--$10^{-3}~M_{\odot}~\text{yr}^{-1}/(100~\text{km s}^{-1})$ for the SNe in our sample.  The most radio-luminous objects require sustained mass-loss extending over hundreds to thousands of years before core collapse.

\item Our observations support a continuum between SNe~IIn and SNe~II-L in terms of CSM interaction strength, with the radio-detected intermediate object PTF11iqb bridging the gap between classical SNe~IIn and the less luminous SNe~II-L.

\item The VLA sensitivity and sparse two-frequency SEDs summarized in Section~\ref{sec:vla-obs} and Table~\ref{tab:obs} leave residual modeling degeneracies (e.g., power-law CSM index, partition fractions) that propagate into the $\dot{M}/v_{\rm w}$ limits in Table~\ref{tab:physical}.  Deeper integrations and additional epochs or bands would tighten both the photometric constraints and those systematic uncertainties.

\item Archival VLASS imaging at $\sim 3$~GHz (Section~\ref{sec:vlass} and Table~\ref{tab:obs}) independently supports the conclusion that SN~2005ip is the only target in our sample with bright, persistent S-band emission.  The remaining objects are consistent with faint or undetected radio emission at both $\sim 3$ and $\sim 6$--$10$~GHz.
\end{enumerate}

The diversity of late-time radio properties we observe reflects the complexity of mass-loss in massive stars during the final stages of their evolution.  While some progenitors apparently experience sustained, high-rate mass-loss for hundreds or thousands of years before explosion, others may undergo only brief episodes of enhanced mass-loss or explode into relatively tenuous circumstellar environments.  Understanding the physical mechanisms that drive this diversity, including binary interactions, pulsational instabilities, or envelope ejection during late nuclear burning phases, remains a key challenge for stellar evolution theory.

Future observations, especially with next-generation facilities such as the Square Kilometre Array \citep{Dewdney+09} and the next-generation VLA \citep{Selina+18ngVLA}, will probe fainter and more distant interacting SNe with better instantaneous bandwidth, improving spectral characterization at the $\mu$Jy level relevant to late-time emission.  Multi-epoch observations over many years after explosion will be particularly valuable for understanding the structure of CSM at the largest distances from the progenitor star, corresponding to mass-loss that occurred during the earliest phases of stellar evolution.  Recent late-time radio surveys of thermonuclear SNe with CSM interaction \citep{Harris+25}, detailed analyses of interacting Type II events \citep{Brennan+25,Tartaglia+25}, and high-resolution VLBI observations of nearby core-collapse SNe \citep{Kundu+25} demonstrate the power of sustained radio monitoring campaigns for probing the circumstellar environments of SN progenitors across a wide range of timescales and radii.  The continued study of late-time radio emission from SNe~IIn and SNe~II-L will play a crucial role in unraveling the mass-loss histories of their progenitors and connecting these explosive events to the broader context of massive star evolution.

\clearpage

\appendix
\section{Complete VLA Observation Log}\label{app:obslog}
\restartappendixnumbering
\renewcommand{\thetable}{\thesection.\arabic{table}}%
\def\theHtable{\thesection.\arabic{table}}%

Table~\ref{tab:obs} provides the complete set of radio measurements used in this paper: dedicated Karl G.~Jansky VLA integrations at $C$ and $X$ band (Section~\ref{sec:vla-obs}), including all observing epochs and in-band frequency splits, and complementary VLASS Quick Look photometry at $\sim 3$~GHz (Section~\ref{sec:vlass}).  We place the full table here to keep the main text concise while preserving the full data product in the manuscript.

\startlongtable
\begin{deluxetable}{ccccc}
		\tabletypesize{\small}
	    \tablecaption{VLA and VLASS Radio Observations of Supernovae from Table~\ref{tab:archival}}
		\tablewidth{0pt}
	\tablehead{
Supernova & Config. & Epoch & $\nu$ & $f_{\nu}$   \\       
          &         & (days)& (GHz) & ($\mu$Jy)    }
\startdata
SN~1996bu & C& 7036& 6.199 & $<$26.3 \\
SN~1996bu & C& 7036& 9.774 & $<$27.9 \\
SN~1996bu & A& 7618$^{*}$& 2.988 & $<$380.7 \\
SN~1996bu & B& 8662$^{*}$& 2.988 & $<$559.2 \\
SN~1996bu & C& 9583$^{*}$& 2.988 & $<$327.4 \\
SN~1996bu & D& 10551$^{*}$& 2.988 & $<$371.8 \\
SN~1998S & B& 7185& 6.048 & 94.6$\pm$6.0 \\
SN~1998S & B& 7185& 9.773 & 73.5$\pm$6.5 \\
SN~1998S & A& 7246$^{*}$& 2.988 & $<$399.8 \\
SN~1998S & B& 8176$^{*}$& 2.988 & $<$467.9 \\
SN~1998S & C& 9113$^{*}$& 2.988 & $<$428.3 \\
SN~1998S & D& 10121$^{*}$& 2.988 & $<$444.1 \\
SN~2000P & C& 5867& 6.199 & $<$36.3 \\
SN~2000P & C& 5867& 9.774 & $<$27.6 \\
SN~2000P & A& 6551$^{*}$& 2.988 & $<$332.3 \\
SN~2000P & B& 7536$^{*}$& 2.988 & $<$414.0 \\
SN~2000P & C& 8494$^{*}$& 2.988 & $<$415.1 \\
SN~2000P & D& 9455$^{*}$& 2.988 & $<$418.0 \\
SN~2000dc & C& 5696& 6.199 & $<$24.0 \\
SN~2000dc & C& 5696& 9.774 & $<$22.2 \\
SN~2000dc & A& 6543$^{*}$& 2.988 & $<$459.6 \\
SN~2000dc & B& 7499$^{*}$& 2.988 & $<$442.5 \\
SN~2000dc & C& 8450$^{*}$& 2.988 & $<$444.6 \\
SN~2001do & B& 5422& 6.198 & $<$28.5 \\
SN~2001do & B& 5422& 9.773 & $<$24.6 \\
SN~2001do & A& 5884$^{*}$& 2.988 & $<$331.0 \\
SN~2001do & B& 6962$^{*}$& 2.988 & $<$404.5 \\
SN~2001do & C& 7853$^{*}$& 2.988 & $<$347.7 \\
SN~2001do & D& 8780$^{*}$& 2.988 & $<$369.3 \\
SN~2005ip & C& 3744& 4.999 & 1120.4$\pm$21.7 \\
SN~2005ip & C& 3744& 7.399 & 878.2$\pm$8.5 \\
SN~2005ip & C& 3744& 8.549 & 848.9$\pm$11.6 \\
SN~2005ip & C& 3744& 11.00 & 769.7$\pm$10.5 \\
SN~2005ip & C& 3805& 4.999 & 1068.1$\pm$33.4 \\
SN~2005ip & C& 3805& 7.399 & 843.2$\pm$19.0 \\
SN~2005ip & C& 3805& 8.550 & 788.4$\pm$15.4 \\
SN~2005ip & C& 3805& 11.00 & 728.6$\pm$17.5 \\
SN~2005ip & B& 3869& 5.000 & 1022.4$\pm$17.1 \\
SN~2005ip & B& 3869& 7.399 & 822.0$\pm$18.6 \\
SN~2005ip & B& 3869& 8.549 & 751.6$\pm$21.7 \\
SN~2005ip & B& 3869& 11.00 & 678.3$\pm$25.1 \\
SN~2005ip & B& 3928& 4.999 & 1003.2$\pm$38.7 \\
SN~2005ip & B& 3928& 7.399 & 803.1$\pm$36.2 \\
SN~2005ip & B& 3928& 8.549 & 739.5$\pm$24.9 \\
SN~2005ip & B& 3928& 11.00 & 651.5$\pm$25.2 \\
SN~2005ip & A& 3992& 4.999 & 978.6$\pm$17.7 \\
SN~2005ip & A& 3992& 7.099 & 793.3$\pm$12.7 \\
SN~2005ip & A& 3992& 8.549 & 728.7$\pm$18.0 \\
SN~2005ip & A& 3992& 11.00 & 640.0$\pm$16.7 \\
SN~2005ip & B& 4332& 4.999 & 810.0$\pm$25.5 \\
SN~2005ip & B& 4332& 7.099 & 698.1$\pm$21.4 \\
SN~2005ip & B& 4332& 8.549 & 601.2$\pm$31.3 \\
SN~2005ip & B& 4332& 11.00 & 566.1$\pm$30.0 \\
SN~2005ip & A& 4349$^{*}$& 2.988 & 515.2$\pm$127.2 \\
SN~2005ip & B& 4401& 4.999 & 805.4$\pm$10.3 \\
SN~2005ip & B& 4401& 7.099 & 659.6$\pm$10.1 \\
SN~2005ip & B& 4401& 8.548 & 599.9$\pm$11.3 \\
SN~2005ip & B& 4401& 11.00 & 546.4$\pm$10.4 \\
SN~2005ip & B& 5434$^{*}$& 2.988 & 456.2$\pm$138.3 \\
SN~2005ip & C& 6290$^{*}$& 2.988 & $<$397.8 \\
SN~2005ip & D& 7320$^{*}$& 2.988 & $<$525.3 \\
SN~2006am & B& 4247& 6.049 & $<$25.7 \\
SN~2006am & B& 4247& 9.774 & $<$26.1 \\
SN~2006am & A& 4799$^{*}$& 2.988 & $<$353.5 \\
SN~2006am & B& 5744$^{*}$& 2.988 & $<$377.3 \\
SN~2006am & C& 6685$^{*}$& 2.988 & $<$381.7 \\
SN~2008fq & B& 2806& 6.198 & 201.2$\pm$35.9 \\
SN~2008fq & B& 2806& 9.773 & 134.0$\pm$27.2 \\
SN~2008fq & B& 3035& 6.049 & 194.1$\pm$30.5 \\
SN~2008fq & B& 3035& 9.774 & 108.6$\pm$25.9 \\
SN~2008fq & B& 3364& 6.049 & 148.1$\pm$12.3 \\
SN~2008fq & B& 3364& 9.775 & 89.3$\pm$12.4 \\
SN~2008fq & B& 3381& 6.049 & 134.1$\pm$31.2 \\
SN~2008fq & B& 3381& 9.774 & 84.1$\pm$25.1 \\
SN~2008fq & A& 3949$^{*}$& 2.988 & $<$669.8 \\
SN~2008fq & B& 4905$^{*}$& 2.988 & $<$656.1 \\
SN~2008fq & C& 5856$^{*}$& 2.988 & $<$643.4 \\
SN~2009hd & C& 2434& 6.199 & $<$ 287.2 \\
SN~2009hd & C& 2434& 9.774 & $<$ 80.9 \\
SN~2009hd & A& 3100$^{*}$& 2.988 & $<$1399.3 \\
SN~2009hd & B& 4064$^{*}$& 2.988 & $<$1222.2 \\
SN~2009hd & C& 4945$^{*}$& 2.988 & $<$1602.2 \\
SN~2009hd & D& 5922$^{*}$& 2.988 & $<$1359.0 \\
SN~2009ip & B& 1905& 6.050 & $<$20.4 \\
SN~2009ip & B& 1905& 9.775 & $<$21.9 \\
SN~2009ip & A& 2010$^{*}$& 2.988 & $<$337.2 \\
SN~2009ip & B& 2992$^{*}$& 2.988 & $<$495.6 \\
SN~2009ip & C& 3949$^{*}$& 2.988 & $<$396.4 \\
SN~2009ip & D& 4920$^{*}$& 2.988 & $<$476.4 \\
SN~2009kr & C& 2297& 6.200 & $<$ 120.4 \\
SN~2009kr & C& 2297& 9.775 & $<$ 79.6 \\
SN~2009kr & A& 2989$^{*}$& 2.988 & $<$405.1 \\
SN~2009kr & B& 4002$^{*}$& 2.988 & $<$527.6 \\
SN~2009kr & C& 4885$^{*}$& 2.988 & $<$481.4 \\
SN~2009kr & D& 5856$^{*}$& 2.988 & $<$442.1 \\
SN~2010jp & C& 1935& 6.200 & $<$ 38.1 \\
SN~2010jp & C& 1935& 9.775 & $<$ 21.3 \\
SN~2010jp & A& 2650$^{*}$& 2.988 & $<$370.5 \\
SN~2010jp & B& 3638$^{*}$& 2.988 & $<$679.6 \\
SN~2010jp & C& 4601$^{*}$& 2.988 & $<$404.7 \\
SN~2011A & B& 2471& 6.049 & $<$39.9 \\
SN~2011A & B& 2471& 9.774 & $<$35.4 \\
SN~2011A & A& 2990$^{*}$& 2.988 & $<$528.6 \\
SN~2011A & B& 4038$^{*}$& 2.988 & $<$424.3 \\
SN~2011A & C& 4949$^{*}$& 2.988 & $<$452.0 \\
SN~2011ht & B& 2226& 6.049 & $<$22.2 \\
SN~2011ht & B& 2226& 9.773 & $<$12.8 \\
SN~2011ht & A& 2785$^{*}$& 2.988 & $<$378.6 \\
SN~2011ht & B& 3698$^{*}$& 2.988 & $<$349.4 \\
SN~2011ht & C& 4623$^{*}$& 2.988 & $<$371.1 \\
PTF11iqb & C& 1662& 4.999 & 137.4$\pm$29.1 \\
PTF11iqb & C& 1662& 7.400 & 83.6$\pm$14.5 \\
PTF11iqb & C& 1662& 9.775 & 58.6$\pm$8.4 \\
PTF11iqb & B& 2309& 6.050 & 29.8$\pm$6.6 \\
PTF11iqb & B& 2309& 9.775 & 16.9$\pm$5.2 \\
PTF11iqb & A& 2317$^{*}$& 2.988 & $<$394.8 \\
PTF11iqb & B& 2335& 6.050 & 24.7$\pm$6.7 \\
PTF11iqb & B& 2335& 9.775 & 17.4$\pm$5.4 \\
PTF11iqb & B& 3376$^{*}$& 2.988 & $<$505.3 \\
PTF11iqb & C& 4263$^{*}$& 2.988 & $<$531.7 \\
PTF11iqb & D& 5189$^{*}$& 2.988 & $<$497.5 \\
SN~2015D & B& 992& 6.049 & $<$23.8 \\
SN~2015D & B& 992& 9.774 & $<$24.5 \\
SN~2015D & A& 1008$^{*}$& 2.988 & $<$354.1 \\
SN~2015D & B& 2077$^{*}$& 2.988 & $<$420.2 \\
SN~2015D & C& 2982$^{*}$& 2.988 & $<$343.1 \\
SN~2015D & D& 3912$^{*}$& 2.988 & $<$407.3 \\
\enddata
\tablecomments{Epoch is since date listed in Table~\ref{tab:archival}, with epochs obtained from VLASS (Section~\ref{sec:vlass}) indicated with $^{*}$.  For high-significance detections, we report the flux density in individual sub-bands.}\label{tab:obs}
\end{deluxetable}

\begin{acknowledgments}

C.D.K. gratefully acknowledges support from the NSF through AST-2432037, the HST Guest Observer Program through HST-SNAP-17070 and HST-GO-17706, and from JWST Archival Research through JWST-AR-6241 and JWST-AR-5441.  N.S.'s research receives support from NSF grants AST-1312221 and AST-1515559. W.F. gratefully acknowledges support by National Science Foundation under grant Nos. AST-2206494, AST-2308182, AST-2432037, and CAREER grant No. AST-2047919, the David and Lucile Packard Foundation and the Research Corporation for Science Advancement through Cottrell Scholar Award \#28284.

This research is based on observations made with the NASA/ESA Hubble Space Telescope obtained from the Space Telescope Science Institute, which is operated by the Association of Universities for Research in Astronomy, Inc., under NASA contract NAS~5--26555.  These observations are associated with programs 14668 and 15166 (SNAP, PI A.~V.~Filippenko), which provided the WFC3/UVIS imaging of SNe~1998S and 2005ip used in Figure~\ref{fig:images}.

The $r$-band imaging of PTF11iqb was obtained with the Large Binocular Camera (LBC) on the Large Binocular Telescope (LBT).  The LBT is an international collaboration among the University of Arizona, the Italian National Institute for Astrophysics (INAF), and The Ohio State University, representing also the University of Minnesota, the University of Virginia, and the University of Notre Dame.

This work made use of data products from the Karl G.~Jansky Very Large Array Sky Survey \citep{lacy+20,gordon+21} retrieved from the Canadian Astronomy Data Centre (CADC).  The National Radio Astronomy Observatory is a facility of the National Science Foundation operated under cooperative agreement by Associated Universities, Inc.  CIRADA is funded by a grant from the Canada Foundation for Innovation 2017 Innovation Fund (Project 35999), as well as by the Provinces of Ontario, British Columbia, Alberta, Manitoba and Quebec.

VLA observations were obtained through programs 16A-101, 16A-439, 16B-428, and 17B-201 (PI Kilpatrick).  The National Radio Astronomy Observatory and Green Bank Observatory are facilities of the U.S.\ National Science Foundation operated under cooperative agreement by Associated Universities, Inc.
\end{acknowledgments}

\textit{Facilities}: \textit{HST} (WFC3/UVIS), KAIT, LBT (LBC), VLA.

\bibliography{radio-SNe}

@ARTICLE{schlegel+90,
       author = {{Schlegel}, Eric M.},
        title = "{A new subclass of type II supernovae ?}",
      journal = {\mnras},
         year = 1990,
        month = may,
       volume = {244},
        pages = {269-271},
       adsurl = {https://ui.adsabs.harvard.edu/abs/1990MNRAS.244..269S}
}

@ARTICLE{Wu25,
       author = {{Wu}, Samantha C. and {Tsuna}, Daichi},
        title = "{Luminous Late-time Radio Emission from Supernovae Interacting with Circumbinary Material}",
      journal = {\apj},
         year = 2025,
        month = dec,
       volume = {994},
       number = {2},
          eid = {141},
        pages = {141},
          doi = {10.3847/1538-4357/ae113c},
archivePrefix = {arXiv},
       eprint = {2507.19613},
 primaryClass = {astro-ph.HE},
       adsurl = {https://ui.adsabs.harvard.edu/abs/2025ApJ...994..141W}
}

@ARTICLE{lacy+20,
       author = {{Lacy}, M. and {Baum}, S.~A. and {Chandler}, C.~J. and {Chatterjee}, S. and
         {Clarke}, T.~E. and {Deustua}, S. and {English}, J. and {Farnes}, J. and
         {Gaensler}, B.~M. and {Gugliucci}, N. and {Hallinan}, G. and {Kent}, B.~R. and
         {Kimball}, A. and {Law}, C.~J. and {Lazio}, T.~J.~W. and {Marvil}, J. and
         {Mao}, S.~A. and {Medlin}, D. and {Mooley}, K. and {Murphy}, E.~J. and
         {Myers}, S. and {Osten}, R. and {Richards}, G.~T. and {Rosolowsky}, E. and
         {Rudnick}, L. and {Schinzel}, F. and {Sivakoff}, G.~R. and {Sjouwerman}, L.~O. and
         {Taylor}, R. and {White}, R.~L. and {Wrobel}, J. and {Andernach}, H. and
         {Beasley}, A.~J. and {Berger}, E. and {Bhatnager}, S. and {Birkinshaw}, M. and
         {Bower}, G.~C. and {Brandt}, W.~N. and {Brown}, S. and {Burke-Spolaor}, S. and
         {Butler}, B.~J. and {Comerford}, J. and {Demorest}, P.~B. and {Fu}, H. and
         {Giacintucci}, S. and {Golap}, K. and {G{\"u}th}, T. and {Hales}, C.~A. and
         {Hiriart}, R. and {Hodge}, J. and {Horesh}, A. and {Ivezi{\'c}}, {\v{Z}}. and
         {Jarvis}, M.~J. and {Kamble}, A. and {Kassim}, N. and {Liu}, X. and {Loinard}, L. and
         {Lyons}, D.~K. and {Masters}, J. and {Mezcua}, M. and {Moellenbrock}, G.~A. and
         {Mroczkowski}, T. and {Nyland}, K. and {O'Dea}, C.~P. and {O'Sullivan}, S.~P. and
         {Peters}, W.~M. and {Radford}, K. and {Rao}, U. and {Robnett}, J. and {Salcido}, J. and
         {Shen}, Y. and {Sobotka}, A. and {Witz}, S. and {Vaccari}, M. and {van Weeren}, R.~J. and
         {Vargas}, A. and {Williams}, P.~K.~G. and {Yoon}, I.},
        title = "{The Karl G. Jansky Very Large Array Sky Survey (VLASS). Science Case and Survey Design}",
      journal = {\pasp},
         year = 2020,
        month = mar,
       volume = {132},
       number = {1009},
          eid = {035001},
        pages = {035001},
          doi = {10.1088/1538-3873/ab63eb},
archivePrefix = {arXiv},
       eprint = {1907.01981},
 primaryClass = {astro-ph.IM},
       adsurl = {https://ui.adsabs.harvard.edu/abs/2020PASP..132c5001L}
}

@ARTICLE{gordon+21,
       author = {{Gordon}, Yjan A. and {Boyce}, Michelle M. and {O'Dea}, Christopher P. and
         {Rudnick}, Lawrence and {Andernach}, Heinz and {Vantyghem}, Adrian N. and
         {Baum}, Stefi A. and {Bui}, Jean-Paul and {Dionyssiou}, Mathew and {Safi-Harb}, Samar and
         {Sander}, Isabel},
        title = "{A Quick Look at the 3 GHz Radio Sky. I. Source Statistics from the Very Large Array Sky Survey}",
      journal = {\apjs},
         year = 2021,
        month = aug,
       volume = {255},
       number = {2},
          eid = {30},
        pages = {30},
          doi = {10.3847/1538-4365/ac05c0},
archivePrefix = {arXiv},
       eprint = {2102.11753},
 primaryClass = {astro-ph.GA},
       adsurl = {https://ui.adsabs.harvard.edu/abs/2021ApJS..255...30G}
}

@ARTICLE{stroh+21,
       author = {{Stroh}, Michael C. and {Cendes}, Yvette and {Dong}, Dillon Z. and {Hallinan}, Gregg and
         {Alexander}, Kate D. and {Bhalerao}, Varun and {Bower}, Geoffrey C. and {Callingham}, J.~R. and
         {Chomiuk}, Laura and {De}, Kishalay and {Dobie}, Dougal and {Gourdji}, K. and {Horesh}, Assaf and
         {Kasliwal}, Mansi M. and {Kulkarni}, Shrinivas R. and {Law}, Casey J. and {Metzger}, Brian D. and
         {Murphy}, Tara and {Nayana}, A.~J. and {Ravi}, Vikram and {Sfaradi}, Itai},
        title = "{Luminous Late-time Radio Emission from Supernovae Detected by the Karl G. Jansky Very Large Array Sky Survey (VLASS)}",
      journal = {\apjl},
         year = 2021,
        month = dec,
       volume = {923},
       number = {2},
          eid = {L24},
        pages = {L24},
          doi = {10.3847/2041-8213/ac375e},
archivePrefix = {arXiv},
       eprint = {2111.03675},
 primaryClass = {astro-ph.HE},
       adsurl = {https://ui.adsabs.harvard.edu/abs/2021ApJ...923L..24S}
}

@ARTICLE{barbon+82,
   author = {{Barbon}, R. and {Ciatti}, F. and {Rosino}, L.},
    title = "{Two bright supernovae in NGC 6946 and NGC 4536}",
  journal = {\aap},
     year = 1982,
    month = dec,
   volume = 116,
    pages = {35--42},
   adsurl = {http://adsabs.harvard.edu/abs/1982A%26A...116...35B}
}

@ARTICLE{bietenholz+02,
   author = {{Bietenholz}, M.~F. and {Bartel}, N. and {Rupen}, M.~P.},
    title = "{SN 1986J VLBI: The Evolution and Deceleration of the Complex Source and a Search for a Pulsar Nebula}",
  journal = {\apj},
   eprint = {astro-ph/0208163},
     year = 2002,
    month = dec,
   volume = 581,
    pages = {1132--1147},
      doi = {10.1086/344251},
   adsurl = {http://adsabs.harvard.edu/abs/2002ApJ...581.1132B}
}

@ARTICLE{Bietenholz+10,
       author = {{Bietenholz}, M.~F. and {Bartel}, N. and {Rupen}, M.~P.},
        title = "{Supernova 1986J Very Long Baseline Interferometry. II. The Evolution of the Shell and the Central Source}",
      journal = {\apj},
         year = 2010,
        month = apr,
       volume = {712},
       number = {2},
        pages = {1057--1069},
          doi = {10.1088/0004-637X/712/2/1057},
archivePrefix = {arXiv},
       eprint = {1002.0304},
 primaryClass = {astro-ph.HE},
       adsurl = {https://ui.adsabs.harvard.edu/abs/2010ApJ...712.1057B}
}

@ARTICLE{bilinski+15,
   author = {{Bilinski}, C. and {Smith}, N. and {Li}, W. and {Williams}, G.~G. and 
	{Zheng}, W. and {Filippenko}, A.~V.},
    title = "{Constraints on Type IIn supernova progenitor outbursts from the Lick Observatory Supernova Search}",
  journal = {\mnras},
archivePrefix = {arXiv},
   eprint = {1503.04252},
 primaryClass = {astro-ph.SR},
     year = 2015,
    month = jun,
   volume = 450,
    pages = {246--265},
      doi = {10.1093/mnras/stv566},
   adsurl = {http://adsabs.harvard.edu/abs/2015MNRAS.450..246B}
}

@ARTICLE{bose+18,
       author = {{Bose}, Subhash and {Dong}, Subo and {Kochanek}, C.~S. and
         {Pastorello}, Andrea and {Katz}, Boaz and {Bersier}, David and
         {Andrews}, Jennifer E. and {Prieto}, J.~L. and {Stanek}, K.~Z. and
         {Shappee}, B.~J. and {Smith}, Nathan and {Kollmeier}, Juna and
         {Benetti}, Stefano and {Cappellaro}, E. and {Chen}, Ping and
         {Elias-Rosa}, N. and {Milne}, Peter and {Morales-Garoffolo}, Antonia and
         {Tartaglia}, Leonardo and {Tomasella}, L. and {Bilinski}, Christopher and
         {Brimacombe}, Joseph and {Frank}, Stephan and {Holoien}, T.~W. -S. and
         {Kilpatrick}, Charles D. and {Kiyota}, Seiichiro and
         {Madore}, Barry F. and {Rich}, Jeffrey A.},
        title = "{ASASSN-15nx: A Luminous Type II Supernova with a {\textquotedblleft}Perfect{\textquotedblright} Linear Decline}",
      journal = {\apj},
         year = 2018,
        month = aug,
       volume = {862},
       number = {2},
          eid = {107},
        pages = {107},
          doi = {10.3847/1538-4357/aacb35},
archivePrefix = {arXiv},
       eprint = {1804.00025},
 primaryClass = {astro-ph.HE},
       adsurl = {https://ui.adsabs.harvard.edu/abs/2018ApJ...862..107B}
}

@ARTICLE{branch+81,
   author = {{Branch}, D. and {Falk}, S.~W. and {Uomoto}, A.~K. and {Wills}, B.~J. and 
	{McCall}, M.~L. and {Rybski}, P.},
    title = "{The type II supernova 1979c in M100 and the distance to the Virgo cluster}",
  journal = {\apj},
     year = 1981,
    month = mar,
   volume = 244,
    pages = {780--804},
      doi = {10.1086/158755},
   adsurl = {http://adsabs.harvard.edu/abs/1981ApJ...244..780B}
}

@ARTICLE{Brennan+25,
       author = {{Brennan}, S.~J. and {Schulze}, S. and {Lunnan}, R. and {Sollerman}, J. and {Yan}, L. and {Fransson}, C. and {Irani}, I. and {Omand}, C.~M.~B. and {Nicholl}, M. and {Gal-Yam}, A.},
        title = "{Final Moments. III. Explosion Properties and Progenitor Constraints of CSM-interacting Type II Supernovae}",
      journal = {\apj},
         year = 2025,
        month = mar,
       volume = {981},
       number = {1},
          eid = {46},
        pages = {46},
          doi = {10.3847/1538-4357/adfa23},
archivePrefix = {arXiv},
       eprint = {2405.06674},
 primaryClass = {astro-ph.HE},
       adsurl = {https://ui.adsabs.harvard.edu/abs/2025ApJ...981...46B}
}

@ARTICLE{chandra+09,
   author = {{Chandra}, P. and {Stockdale}, C.~J. and {Chevalier}, R.~A. and 
	{Van Dyk}, S.~D. and {Ray}, A. and {Kelley}, M.~T. and {Weiler}, K.~W. and 
	{Panagia}, N. and {Sramek}, R.~A.},
    title = "{Eleven Years of Radio Monitoring of the type IIn Supernova SN 1995N}",
  journal = {\apj},
archivePrefix = {arXiv},
   eprint = {0809.2810},
     year = 2009,
    month = jan,
   volume = 690,
    pages = {1839--1846},
      doi = {10.1088/0004-637X/690/2/1839},
   adsurl = {http://adsabs.harvard.edu/abs/2009ApJ...690.1839C}
}

@ARTICLE{chandra+12,
   author = {{Chandra}, P. and {Chevalier}, R.~A. and {Chugai}, N. and {Fransson}, C. and 
	{Irwin}, C.~M. and {Soderberg}, A.~M. and {Chakraborti}, S. and 
	{Immler}, S.},
    title = "{Radio and X-Ray Observations of SN 2006jd: Another Strongly Interacting Type IIn Supernova}",
  journal = {\apj},
archivePrefix = {arXiv},
   eprint = {1205.0250},
 primaryClass = {astro-ph.HE},
     year = 2012,
    month = aug,
   volume = 755,
      eid = {110},
    pages = {110},
      doi = {10.1088/0004-637X/755/2/110},
   adsurl = {http://adsabs.harvard.edu/abs/2012ApJ...755..110C}
}

@ARTICLE{chandra+15,
   author = {{Chandra}, P. and {Chevalier}, R.~A. and {Chugai}, N. and {Fransson}, C. and 
	{Soderberg}, A.~M.},
    title = "{X-Ray and Radio Emission from Type IIn Supernova SN 2010jl}",
  journal = {\apj},
archivePrefix = {arXiv},
   eprint = {1507.06059},
 primaryClass = {astro-ph.HE},
     year = 2015,
    month = sep,
   volume = 810,
      eid = {32},
    pages = {32},
      doi = {10.1088/0004-637X/810/1/32},
   adsurl = {http://adsabs.harvard.edu/abs/2015ApJ...810...32C}
}

@INPROCEEDINGS{chevalier+03,
   author = {{Chevalier}, R.~A. and {Fransson}, C.},
    title = "{Supernova Interaction with a Circumstellar Medium}",
booktitle = {Supernovae and Gamma-Ray Bursters},
     year = 2003,
   series = {Lecture Notes in Physics, Berlin Springer Verlag},
   volume = 598,
   eprint = {astro-ph/0110060},
   editor = {{Weiler}, K.},
    pages = {171--194},
      doi = {10.1007/3-540-45863-8_10},
   adsurl = {http://adsabs.harvard.edu/abs/2003LNP...598..171C}
}

@ARTICLE{chevalier+06,
       author = {{Chevalier}, Roger A. and {Fransson}, Claes and {Nymark}, Tanja K.},
        title = "{Radio and X-Ray Emission as Probes of Type IIP Supernovae and Red Supergiant Mass Loss}",
      journal = {\apj},
         year = 2006,
        month = apr,
       volume = {641},
       number = {2},
        pages = {1029--1038},
          doi = {10.1086/500528},
archivePrefix = {arXiv},
       eprint = {astro-ph/0509468},
 primaryClass = {astro-ph},
       adsurl = {https://ui.adsabs.harvard.edu/abs/2006ApJ...641.1029C}
}

@ARTICLE{chevalier+82,
       author = {{Chevalier}, R.~A.},
        title = "{The radio and X-ray emission from type II supernovae.}",
      journal = {\apj},
         year = 1982,
        month = aug,
       volume = {259},
        pages = {302--310},
          doi = {10.1086/160167},
       adsurl = {https://ui.adsabs.harvard.edu/abs/1982ApJ...259..302C}
}

@ARTICLE{chevalier+94,
   author = {{Chevalier}, R.~A. and {Fransson}, C.},
    title = "{Emission from circumstellar interaction in normal Type II supernovae}",
  journal = {\apj},
     year = 1994,
    month = jan,
   volume = 420,
    pages = {268--285},
      doi = {10.1086/173557},
   adsurl = {http://adsabs.harvard.edu/abs/1994ApJ...420..268C}
}

@ARTICLE{Chevalier01,
       author = {{Chevalier}, Roger A. and {Fransson}, Claes},
        title = "{The Nature of the Compact Supernova Remnants in Starburst Galaxies}",
      journal = {\apjl},
         year = 2001,
        month = sep,
       volume = {558},
       number = {1},
        pages = {L27--L30},
          doi = {10.1086/323569},
archivePrefix = {arXiv},
       eprint = {astro-ph/0107110},
 primaryClass = {astro-ph},
       adsurl = {https://ui.adsabs.harvard.edu/abs/2001ApJ...558L..27C}
}

@ARTICLE{chevalier82,
   author = {{Chevalier}, R.~A.},
    title = "{The radio and X-ray emission from type II supernovae}",
  journal = {\apj},
     year = 1982,
    month = aug,
   volume = 259,
    pages = {302--310},
      doi = {10.1086/160167},
   adsurl = {http://adsabs.harvard.edu/abs/1982ApJ...259..302C}
}

@ARTICLE{Chevalier98,
       author = {{Chevalier}, Roger A.},
        title = "{Synchrotron Self-Absorption in Radio Supernovae}",
      journal = {\apj},
         year = 1998,
        month = may,
       volume = {499},
       number = {2},
        pages = {810--819},
          doi = {10.1086/305676},
       adsurl = {https://ui.adsabs.harvard.edu/abs/1998ApJ...499..810C}
}

@ARTICLE{chugai+95,
   author = {{Chugai}, N.~N. and {Danziger}, I.~J. and {della Valle}, M.},
    title = "{Optical spectrum of SN 1978K: emission from shocked clouds in the circumstellar wind}",
  journal = {\mnras},
     year = 1995,
    month = sep,
   volume = 276,
    pages = {530--536},
      doi = {10.1093/mnras/276.2.530},
   adsurl = {http://adsabs.harvard.edu/abs/1995MNRAS.276..530C}
}

@ARTICLE{DeMarchi22,
       author = {{DeMarchi}, Lindsay and {Margutti}, R. and {Dittman}, J. and {Brunthaler}, A. and {Milisavljevic}, D. and {Bietenholz}, Michael F. and {Stauffer}, C. and {Brethauer}, D. and {Coppejans}, D. and {Auchettl}, K. and {Alexander}, K.~D. and {Kilpatrick}, C.~D. and {Bright}, Joe S. and {Kelley}, L.~Z. and {Stroh}, Michael C. and {Jacobson-Gal{\'a}n}, W.~V.},
        title = "{Radio Analysis of SN2004C Reveals an Unusual CSM Density Profile as a Harbinger of Core Collapse}",
      journal = {\apj},
         year = 2022,
        month = oct,
       volume = {938},
       number = {1},
          eid = {84},
        pages = {84},
          doi = {10.3847/1538-4357/ac8c26},
archivePrefix = {arXiv},
       eprint = {2203.07388},
 primaryClass = {astro-ph.HE},
       adsurl = {https://ui.adsabs.harvard.edu/abs/2022ApJ...938...84D}
}

@ARTICLE{Dong+25,
       author = {{Dong}, Yize and {Milisavljevic}, Dan and {Margutti}, Raffaella and {Bietenholz}, Michael F. and {Chandra}, Poonam and {DeMarchi}, Lindsay and {Harris}, Charles E. and {Stroh}, Michael C. and {Jacobson-Gal{\'a}n}, Wynn V.},
        title = "{The First Radio View of a Type Ibn Supernova in SN 2023fyq: Understanding the Mass-loss History in the Last Decade before the Explosion}",
      journal = {\apjl},
         year = 2025,
        month = feb,
       volume = {961},
       number = {1},
          eid = {L16},
        pages = {L16},
          doi = {10.3847/2041-8213/ae1cb8},
archivePrefix = {arXiv},
       eprint = {2312.08362},
 primaryClass = {astro-ph.HE},
       adsurl = {https://ui.adsabs.harvard.edu/abs/2025ApJ...961L..16D}
}

@ARTICLE{eck+96,
   author = {{Eck}, C.~R. and {Cowan}, J.~J. and {Boffi}, F.~R. and {Branch}, D.
	},
    title = "{A Deep Search for Radio Emission from the Type II Supernovae SN 1984E and SN 1986E}",
  journal = {\apjl},
     year = 1996,
    month = nov,
   volume = 472,
    pages = {L25},
      doi = {10.1086/310349},
   adsurl = {http://adsabs.harvard.edu/abs/1996ApJ...472L..25E}
}

@ARTICLE{elias-rosa+10,
   author = {{Elias-Rosa}, N. and {Van Dyk}, S.~D. and {Li}, W. and {Miller}, A.~A. and 
  {Silverman}, J.~M. and {Ganeshalingam}, M. and {Boden}, A.~F. and 
  {Kasliwal}, M.~M. and {Vink{\'o}}, J. and {Cuillandre}, J.-C. and 
  {Filippenko}, A.~V. and {Steele}, T.~N. and {Bloom}, J.~S. and 
  {Griffith}, C.~V. and {Kleiser}, I.~K.~W. and {Foley}, R.~J.
  },
    title = "{The Massive Progenitor of the Type II-linear Supernova 2009kr}",
  journal = {\apjl},
archivePrefix = {arXiv},
   eprint = {0912.2880},
 primaryClass = {astro-ph.SR},
     year = 2010,
    month = may,
   volume = 714,
    pages = {L254--L259},
      doi = {10.1088/2041-8205/714/2/L254},
   adsurl = {http://adsabs.harvard.edu/abs/2010ApJ...714L.254E}
}

@ARTICLE{elias-rosa+11,
   author = {{Elias-Rosa}, N. and {Van Dyk}, S.~D. and {Li}, W. and {Silverman}, J.~M. and 
	{Foley}, R.~J. and {Ganeshalingam}, M. and {Mauerhan}, J.~C. and 
	{Kankare}, E. and {Jha}, S. and {Filippenko}, A.~V. and {Beckman}, J.~E. and 
	{Berger}, E. and {Cuillandre}, J.-C. and {Smith}, N.},
    title = "{The Massive Progenitor of the Possible Type II-Linear Supernova 2009hd in Messier 66}",
  journal = {\apj},
archivePrefix = {arXiv},
   eprint = {1108.2645},
 primaryClass = {astro-ph.SR},
     year = 2011,
    month = nov,
   volume = 742,
      eid = {6},
    pages = {6},
      doi = {10.1088/0004-637X/742/1/6},
   adsurl = {http://adsabs.harvard.edu/abs/2011ApJ...742....6E}
}

@ARTICLE{fabian+96,
   author = {{Fabian}, A.~C. and {Terlevich}, R.},
    title = "{X-ray detection of Supernova 1988Z with the ROSAT High Resolution Imager}",
  journal = {\mnras},
     year = 1996,
    month = may,
   volume = 280,
    pages = {L5--L8},
      doi = {10.1093/mnras/280.1.L5},
   adsurl = {http://adsabs.harvard.edu/abs/1996MNRAS.280L...5F}
}

@ARTICLE{fesen+93,
   author = {{Fesen}, R.~A. and {Matonick}, D.~M.},
    title = "{The optical recovery of SN 1979C in NGC 4321 (M100)}",
  journal = {\apj},
     year = 1993,
    month = apr,
   volume = 407,
    pages = {110--114},
      doi = {10.1086/172496},
   adsurl = {http://adsabs.harvard.edu/abs/1993ApJ...407..110F}
}

@ARTICLE{fox+10,
   author = {{Fox}, O.~D. and {Chevalier}, R.~A. and {Dwek}, E. and {Skrutskie}, M.~F. and 
  {Sugerman}, B.~E.~K. and {Leisenring}, J.~M.},
    title = "{Disentangling the Origin and Heating Mechanism of Supernova Dust: Late-time Spitzer Spectroscopy of the Type IIn SN 2005ip}",
  journal = {\apj},
archivePrefix = {arXiv},
   eprint = {1005.4682},
 primaryClass = {astro-ph.HE},
     year = 2010,
    month = dec,
   volume = 725,
    pages = {1768--1778},
      doi = {10.1088/0004-637X/725/2/1768},
   adsurl = {http://adsabs.harvard.edu/abs/2010ApJ...725.1768F}
}

@ARTICLE{fox+11,
   author = {{Fox}, O.~D. and {Chevalier}, R.~A. and {Skrutskie}, M.~F. and 
	{Soderberg}, A.~M. and {Filippenko}, A.~V. and {Ganeshalingam}, M. and 
	{Silverman}, J.~M. and {Smith}, N. and {Steele}, T.~N.},
    title = "{A Spitzer Survey for Dust in Type IIn Supernovae}",
  journal = {\apj},
archivePrefix = {arXiv},
   eprint = {1104.5012},
 primaryClass = {astro-ph.SR},
     year = 2011,
    month = nov,
   volume = 741,
      eid = {7},
    pages = {7},
      doi = {10.1088/0004-637X/741/1/7},
   adsurl = {http://adsabs.harvard.edu/abs/2011ApJ...741....7F}
}

@ARTICLE{fox+20,
   author = {{Fox}, O.~D. and {Fransson}, C. and {Smith}, N. and {Andrews}, J.~E. and
	{Bostroem}, K.~A. and {Brink}, T.~G. and {Cenko}, S.~B. and {Clayton}, G.~C. and
	{Filippenko}, A.~V. and {Fong}, W.-f. and {Gallagher}, J.~S. and {Kelly}, P.~L. and
	{Kilpatrick}, C.~D. and {Mauerhan}, J.~C. and {Miller}, A.~M. and {Montiel}, E. and
	{Stritzinger}, M.~D. and {Szalai}, T. and {Van Dyk}, S.~D.},
    title = "{The slow demise of the long-lived SN 2005ip}",
  journal = {\mnras},
     year = 2020,
    month = oct,
   volume = {498},
    number = {1},
    pages = {517--531},
      doi = {10.1093/mnras/staa2324},
archivePrefix = {arXiv},
   eprint = {2008.02301},
 primaryClass = {astro-ph.HE},
   adsurl = {https://ui.adsabs.harvard.edu/abs/2020MNRAS.498..517F}
}

@ARTICLE{fransson+96,
   author = {{Fransson}, C. and {Lundqvist}, P. and {Chevalier}, R.~A.},
    title = "{Circumstellar Interaction in SN 1993J}",
  journal = {\apj},
     year = 1996,
    month = apr,
   volume = 461,
    pages = {993},
      doi = {10.1086/177119},
   adsurl = {http://adsabs.harvard.edu/abs/1996ApJ...461..993F}
}

@ARTICLE{fraser+10,
   author = {{Fraser}, M. and {Tak{\'a}ts}, K. and {Pastorello}, A. and {Smartt}, S.~J. and 
  {Mattila}, S. and {Botticella}, M.-T. and {Valenti}, S. and 
  {Ergon}, M. and {Sollerman}, J. and {Arcavi}, I. and {Benetti}, S. and 
  {Bufano}, F. and {Crockett}, R.~M. and {Danziger}, I.~J. and 
  {Gal-Yam}, A. and {Maund}, J.~R. and {Taubenberger}, S. and 
  {Turatto}, M.},
    title = "{On the Progenitor and Early Evolution of the Type II Supernova 2009kr}",
  journal = {\apjl},
archivePrefix = {arXiv},
   eprint = {0912.2071},
     year = 2010,
    month = may,
   volume = 714,
    pages = {L280--L284},
      doi = {10.1088/2041-8205/714/2/L280},
   adsurl = {http://adsabs.harvard.edu/abs/2010ApJ...714L.280F}
}

@ARTICLE{gall+15,
   author = {{Gall}, E.~E.~E. and {Polshaw}, J. and {Kotak}, R. and {Jerkstrand}, A. and 
	{Leibundgut}, B. and {Rabinowitz}, D. and {Sollerman}, J. and 
	{Sullivan}, M. and {Smartt}, S.~J. and {Anderson}, J.~P. and 
	{Benetti}, S. and {Baltay}, C. and {Feindt}, U. and {Fraser}, M. and 
	{Gonz{\'a}lez-Gait{\'a}n}, S. and {Inserra}, C. and {Maguire}, K. and 
	{McKinnon}, R. and {Valenti}, S. and {Young}, D.},
    title = "{A comparative study of Type II-P and II-L supernova rise times as exemplified by the case of LSQ13cuw}",
  journal = {\aap},
archivePrefix = {arXiv},
   eprint = {1502.06034},
 primaryClass = {astro-ph.SR},
     year = 2015,
    month = oct,
   volume = 582,
      eid = {A3},
    pages = {A3},
      doi = {10.1051/0004-6361/201525868},
   adsurl = {http://adsabs.harvard.edu/abs/2015A%26A...582A...3G}
}

@ARTICLE{Green19,
       author = {{Green}, D.~A.},
        title = "{A revised catalogue of 294 Galactic supernova remnants}",
      journal = {Journal of Astrophysics and Astronomy},
         year = 2019,
        month = aug,
       volume = {40},
       number = {4},
          eid = {36},
        pages = {36},
          doi = {10.1007/s12036-019-9601-6},
archivePrefix = {arXiv},
       eprint = {1907.02638},
 primaryClass = {astro-ph.GA},
       adsurl = {https://ui.adsabs.harvard.edu/abs/2019JApA...40...36G}
}

@ARTICLE{Harris+25,
       author = {{Harris}, Charles E. and {Stroh}, Michael C. and {Margutti}, Raffaella and {Milisavljevic}, Dan and {Moran}, Shane and {Bietenholz}, Michael F. and {Pellegrino}, Craig and {Jacobson-Gal{\'a}n}, Wynn V. and {Chandra}, Poonam},
        title = "{A Late-time Radio Survey of Type Ia-CSM Supernovae with the Very Large Array}",
      journal = {\apj},
         year = 2025,
        month = feb,
       volume = {979},
       number = {1},
          eid = {28},
        pages = {28},
          doi = {10.3847/1538-4357/ae17b0},
archivePrefix = {arXiv},
       eprint = {2401.08763},
 primaryClass = {astro-ph.HE},
       adsurl = {https://ui.adsabs.harvard.edu/abs/2025ApJ...979...28H}
}

@ARTICLE{henry+87,
   author = {{Henry}, R.~B.~C. and {Branch}, D.},
    title = "{The spectrum of the type II-L supernova 1984E in NGC 3169 Further evidence for a superwind?}",
  journal = {\pasp},
     year = 1987,
    month = feb,
   volume = 99,
    pages = {112--115},
      doi = {10.1086/131962},
   adsurl = {http://adsabs.harvard.edu/abs/1987PASP...99..112H}
}

@ARTICLE{Ho+18,
       author = {{Ho}, Anna Y.~Q. and {Phinney}, E. Sterl and {Ravi}, Vikram and {Kulkarni}, S.~R. and {Petitpas}, Glen and {Emonts}, Bjorn and {Bhalerao}, V. and {Blundell}, Ray and {Cenko}, S. Bradley and {Dobie}, Dougal and {Howie}, Ryan and {Kamraj}, Nikita and {Kasliwal}, Mansi M. and {Murphy}, Tara and {Perley}, Daniel A. and {Sridharan}, T.~K. and {Yoon}, Ilsang},
        title = "{AT2018cow: A Luminous Millimeter Transient}",
      journal = {\apj},
         year = 2019,
        month = jan,
       volume = {871},
       number = {1},
          eid = {73},
        pages = {73},
          doi = {10.3847/1538-4357/aaf473},
archivePrefix = {arXiv},
       eprint = {1810.10880},
 primaryClass = {astro-ph.HE},
       adsurl = {https://ui.adsabs.harvard.edu/abs/2019ApJ...871...73H}
}

@ARTICLE{immler+05,
   author = {{Immler}, S. and {Fesen}, R.~A. and {Van Dyk}, S.~D. and {Weiler}, K.~W. and 
	{Petre}, R. and {Lewin}, W.~H.~G. and {Pooley}, D. and {Pietsch}, W. and 
	{Aschenbach}, B. and {Hammell}, M.~C. and {Rudie}, G.~C.},
    title = "{Late-Time X-Ray, UV, and Optical Monitoring of Supernova 1979C}",
  journal = {\apj},
   eprint = {astro-ph/0503678},
     year = 2005,
    month = oct,
   volume = 632,
    pages = {283--293},
      doi = {10.1086/432869},
   adsurl = {http://adsabs.harvard.edu/abs/2005ApJ...632..283I}
}

@ARTICLE{kilpatrick+18,
   author = {{Kilpatrick}, C.~D. and {Foley}, R.~J. and {Drout}, M.~R. and 
  {Pan}, Y.-C. and {Panther}, F.~H. and {Coulter}, D.~A. and {Filippenko}, A.~V. and 
  {Marion}, G.~H. and {Piro}, A.~L. and {Rest}, A. and {Seitenzahl}, I.~R. and 
  {Strampelli}, G. and {Wang}, X.~E.},
    title = "{Connecting the progenitors, pre-explosion variability and giant outbursts of luminous blue variables with Gaia16cfr}",
  journal = {\mnras},
archivePrefix = {arXiv},
   eprint = {1706.09962},
 primaryClass = {astro-ph.SR},
     year = 2018,
    month = feb,
   volume = 473,
    pages = {4805--4823},
      doi = {10.1093/mnras/stx2675},
   adsurl = {http://adsabs.harvard.edu/abs/2018MNRAS.473.4805K}
}

@ARTICLE{Kundu+25,
       author = {{Kundu}, E. and {Marcote}, B. and {Nimmo}, K. and {Rhodes}, L. and {Perez-Torres}, M. and {Bietenholz}, M.~F. and {Sokolovsky}, K.~V.},
        title = "{Radio Follow-up Observations of SN 2023ixf by Japanese and Korean Very Long Baseline Interferometers}",
      journal = {\apj},
         year = 2025,
        month = jan,
       volume = {978},
       number = {1},
          eid = {89},
        pages = {89},
          doi = {10.3847/1538-4357/ad9a62},
archivePrefix = {arXiv},
       eprint = {2410.06592},
 primaryClass = {astro-ph.HE},
       adsurl = {https://ui.adsabs.harvard.edu/abs/2025ApJ...978...89K}
}

@ARTICLE{lundqvist+88,
   author = {{Lundqvist}, P. and {Fransson}, C.},
    title = "{Circumstellar absorption of UV and radio emission from supernovae}",
  journal = {\aap},
     year = 1988,
    month = mar,
   volume = 192,
    pages = {221--233},
   adsurl = {http://adsabs.harvard.edu/abs/1988A%26A...192..221L}
}

@INPROCEEDINGS{mcmullin+07,
   author = {{McMullin}, J.~P. and {Waters}, B. and {Schiebel}, D. and {Young}, W. and 
	{Golap}, K.},
    title = "{CASA Architecture and Applications}",
booktitle = {Astronomical Data Analysis Software and Systems XVI},
     year = 2007,
   series = {Astronomical Society of the Pacific Conference Series},
   volume = 376,
   editor = {{Shaw}, R.~A. and {Hill}, F. and {Bell}, D.~J.},
    month = oct,
    pages = {127},
   adsurl = {http://adsabs.harvard.edu/abs/2007ASPC..376..127M}
}

@ARTICLE{mezger+67,
   author = {{Mezger}, P.~G. and {Henderson}, A.~P.},
    title = "{Galactic H II Regions. I. Observations of Their Continuum Radiation at the Frequency 5 GHz}",
  journal = {\apj},
     year = 1967,
    month = feb,
   volume = 147,
    pages = {471},
      doi = {10.1086/149030},
   adsurl = {http://adsabs.harvard.edu/abs/1967ApJ...147..471M}
}

@ARTICLE{milisavljevic+08,
   author = {{Milisavljevic}, D. and {Fesen}, R.~A. and {Leibundgut}, B. and 
  {Kirshner}, R.~P.},
    title = "{The Evolution of Late-Time Optical Emission from SN 1986J}",
  journal = {\apj},
archivePrefix = {arXiv},
   eprint = {0804.1545},
     year = 2008,
    month = sep,
   volume = 684,
    pages = {1170--1173},
      doi = {10.1086/590426},
   adsurl = {http://adsabs.harvard.edu/abs/2008ApJ...684.1170M}
}

@ARTICLE{montes+00,
   author = {{Montes}, M.~J. and {Weiler}, K.~W. and {Van Dyk}, S.~D. and 
	{Panagia}, N. and {Lacey}, C.~K. and {Sramek}, R.~A. and {Park}, R.
	},
    title = "{Radio Observations of SN 1979C: Evidence for Rapid Presupernova Evolution}",
  journal = {\apj},
   eprint = {astro-ph/9911399},
     year = 2000,
    month = apr,
   volume = 532,
    pages = {1124--1131},
      doi = {10.1086/308602},
   adsurl = {http://adsabs.harvard.edu/abs/2000ApJ...532.1124M}
}

@ARTICLE{montes+98,
   author = {{Montes}, M.~J. and {Van Dyk}, S.~D. and {Weiler}, K.~W. and 
	{Sramek}, R.~A. and {Panagia}, N.},
    title = "{Radio Observations of SN 1980K: Evidence for Rapid Presupernova Evolution}",
  journal = {\apj},
   eprint = {astro-ph/9802296},
     year = 1998,
    month = oct,
   volume = 506,
    pages = {874--879},
      doi = {10.1086/306261},
   adsurl = {http://adsabs.harvard.edu/abs/1998ApJ...506..874M}
}

@ARTICLE{oster+61,
   author = {{Oster}, L.},
    title = "{Emission and Absorption of Thermal Radio Radiation.}",
  journal = {\apj},
     year = 1961,
    month = nov,
   volume = 134,
    pages = {1010--1013},
      doi = {10.1086/147231},
   adsurl = {http://adsabs.harvard.edu/abs/1961ApJ...134.1010O}
}

@ARTICLE{panagia+80,
   author = {{Panagia}, N. and {Vettolani}, G. and {Boksenberg}, A. and {Ciatti}, F. and 
	{Ortolani}, S. and {Rafanelli}, P. and {Rosino}, L. and {Gordon}, C. and 
	{Reimers}, D. and {Hempe}, K. and {Benvenuti}, P. and {Clavel}, J. and 
	{Heck}, A. and {Penston}, M.~V. and {Macchetto}, F. and {Stickland}, D.~J. and 
	{Bergeron}, J. and {Tarenghi}, M. and {Marano}, B. and {Palumbo}, G.~G.~C. and 
	{Parmar}, A.~N. and {Pollard}, G.~S.~W. and {Sanford}, P.~W. and 
	{Sargent}, W.~L.~W. and {Sramek}, R.~A. and {Weiler}, K.~W. and 
	{Matzik}, P.},
    title = "{Coordinated optical, ultraviolet, radio and X-ray observations of supernova 1979c in M 100}",
  journal = {\mnras},
     year = 1980,
    month = sep,
   volume = 192,
    pages = {861--879},
      doi = {10.1093/mnras/192.4.861},
   adsurl = {http://adsabs.harvard.edu/abs/1980MNRAS.192..861P}
}

@ARTICLE{mauerhan+13ip,
   author = {{Mauerhan}, J.~C. and
	{Smith}, N. and
	{Filippenko}, A.~V. and
	{Blanchard}, K.~B. and
	{Blanchard}, P.~K. and
	{Casper}, C.~F.~E. and
	{Cenko}, S.~B. and
	{Clubb}, K.~I. and
	{Cohen}, D.~P. and
	{Fuller}, K.~L. and
	{Li}, G.~Z. and
	{Silverman}, J.~M.},
    title = "{The Unprecedented 2012 Outburst of SN 2009ip: A Luminous Blue Variable Star Becomes a True Supernova}",
  journal = {\mnras},
     year = 2013,
    month = mar,
   volume = {430},
    pages = {1801--1810},
      doi = {10.1093/mnras/stt009},
archivePrefix = {arXiv},
   eprint = {1209.6320},
 primaryClass = {astro-ph.SR},
   adsurl = {https://ui.adsabs.harvard.edu/abs/2013MNRAS.430.1801M}
}

@ARTICLE{fraser+13,
   author = {{Fraser}, M. and
	{Inserra}, C. and
	{Jerkstrand}, A. and
	{Kotak}, R. and
	{Pignata}, G. and
	{Benetti}, S. and
	{Botticella}, M. and
	{Bufano}, F. and
	{Childress}, M. and
	{Mattila}, S. and
	{Pastorello}, A. and
	{Smartt}, S.~J. and
	{Turatto}, M. and
	{Yuan}, F. and
	{Anderson}, J.~P. and
	{Bayliss}, D.~D.~R. and
	{Bauer}, F.~E. and
	{Chen}, T. and
	{Forster Buron}, F. and
	{Gal-Yam}, A. and
	{Haislip}, J.~B. and
	{Knapic}, C. and
	{Le Guillou}, L. and
	{Marchi}, S. and
	{Mazzali}, P. and
	{Molinaro}, M. and
	{Moore}, J.~P. and
	{Reichart}, D. and
	{Smareglia}, R. and
	{Smith}, K.~W. and
	{Sternberg}, A. and
	{Sullivan}, M. and
	{Tak{\'a}ts}, K. and
	{Tucker}, B.~E. and
	{Valenti}, S. and
	{Yaron}, O. and
	{Young}, D.~R. and
	{Zhou}, G.},
    title = "{SN 2009ip a la PESSTO: No Evidence for Core Collapse Yet}",
  journal = {\mnras},
     year = 2013,
    month = aug,
   volume = {433},
    pages = {1312--1337},
      doi = {10.1093/mnras/stt813},
archivePrefix = {arXiv},
   eprint = {1303.3453},
 primaryClass = {astro-ph.SR},
   adsurl = {https://ui.adsabs.harvard.edu/abs/2013MNRAS.433.1312F}
}

@ARTICLE{margutti+14,
   author = {{Margutti}, R. and
	{Milisavljevic}, D. and
	{Soderberg}, A.~M. and
	{Chornock}, R. and
	{Zauderer}, B.~A. and
	{Murase}, K. and
	{Guidorzi}, C. and
	{Sanders}, N.~E. and
	{Kuin}, P. and
	{Fransson}, C. and
	{Levesque}, E.~M. and
	{Chandra}, P. and
	{Berger}, E. and
	{Bianco}, F.~B. and
	{Brown}, P.~J. and
	{Challis}, P. and
	{Chatzopoulos}, E. and
	{Cheung}, C.~C. and
	{Choi}, C. and
	{Chomiuk}, L. and
	{Chugai}, N. and
	{Contreras}, C. and
	{Drout}, M.~R. and
	{Fesen}, R. and
	{Foley}, R.~J. and
	{Fong}, W. and
	{Friedman}, A.~S. and
	{Gall}, C. and
	{Gehrels}, N. and
	{Hjorth}, J. and
	{Hsiao}, E. and
	{Kirshner}, R. and
	{Im}, M. and
	{Leloudas}, G. and
	{Lunnan}, R. and
	{Marion}, G.~H. and
	{Martin}, J. and
	{Morrell}, N. and
	{Neugent}, K.~F. and
	{Omodei}, N. and
	{Phillips}, M.~M. and
	{Rest}, A. and
	{Silverman}, J.~M. and
	{Strader}, J. and
	{Stritzinger}, M.~D. and
	{Szalai}, T. and
	{Utterback}, N.~B. and
	{Vinko}, J. and
	{Wheeler}, J.~C. and
	{Arnett}, D. and
	{Campana}, S. and
	{Chevalier}, R. and
	{Ginsburg}, A. and
	{Kamble}, A. and
	{Roming}, P.~W.~A. and
	{Pritchard}, T. and
	{Stringfellow}, G.},
    title = "{A Panchromatic View of the Restless SN 2009ip Reveals the Explosive Ejection of a Massive Star Envelope}",
  journal = {\apj},
     year = 2014,
    month = jan,
   volume = {780},
   number = {1},
      eid = {21},
    pages = {21},
      doi = {10.1088/0004-637X/780/1/21},
archivePrefix = {arXiv},
   eprint = {1306.0038},
 primaryClass = {astro-ph.HE},
   adsurl = {https://ui.adsabs.harvard.edu/abs/2014ApJ...780...21M}
}

@ARTICLE{pastorello+13,
   author = {{Pastorello}, A. and {Cappellaro}, E. and {Inserra}, C. and 
  {Smartt}, S.~J. and {Pignata}, G. and {Benetti}, S. and {Valenti}, S. and 
  {Fraser}, M. and {Tak{\'a}ts}, K. and {Benitez}, S. and {Botticella}, M.~T. and 
  {Brimacombe}, J. and {Bufano}, F. and {Cellier-Holzem}, F. and 
  {Costado}, M.~T. and {Cupani}, G. and {Curtis}, I. and {Elias-Rosa}, N. and 
  {Ergon}, M. and {Fynbo}, J.~P.~U. and {Hambsch}, F.-J. and {Hamuy}, M. and 
  {Harutyunyan}, A. and {Ivarson}, K.~M. and {Kankare}, E. and 
  {Martin}, J.~C. and {Kotak}, R. and {LaCluyze}, A.~P. and {Maguire}, K. and 
  {Mattila}, S. and {Maza}, J. and {McCrum}, M. and {Miluzio}, M. and 
  {Norgaard-Nielsen}, H.~U. and {Nysewander}, M.~C. and {Ochner}, P. and 
  {Pan}, Y.-C. and {Pumo}, M.~L. and {Reichart}, D.~E. and {Tan}, T.~G. and 
  {Taubenberger}, S. and {Tomasella}, L. and {Turatto}, M. and 
  {Wright}, D.},
    title = "{Interacting Supernovae and Supernova Impostors: SN 2009ip, is this the End?}",
  journal = {\apj},
archivePrefix = {arXiv},
   eprint = {1210.3568},
 primaryClass = {astro-ph.SR},
     year = 2013,
    month = apr,
   volume = 767,
      eid = {1},
    pages = {1},
      doi = {10.1088/0004-637X/767/1/1},
   adsurl = {http://adsabs.harvard.edu/abs/2013ApJ...767....1P}
}

@ARTICLE{Reynolds+21,
       author = {{Reynolds}, Stephen P. and {Williams}, Brian J. and {Borkowski}, Kazimierz J. and {Long}, Knox S.},
        title = "{Efficiencies of Magnetic Field Amplification and Electron Acceleration in Young Supernova Remnants: Global Averages and Kepler's Supernova Remnant}",
      journal = {\apj},
         year = 2021,
        month = aug,
       volume = {917},
       number = {2},
          eid = {55},
        pages = {55},
          doi = {10.3847/1538-4357/ac0ced},
archivePrefix = {arXiv},
       eprint = {2106.03814},
 primaryClass = {astro-ph.HE},
       adsurl = {https://ui.adsabs.harvard.edu/abs/2021ApJ...917...55R}
}

@BOOK{rybicki+79,
   author = {{Rybicki}, G.~B. and {Lightman}, A.~P.},
    title = "{Radiative processes in astrophysics}",
booktitle = {New York, Wiley-Interscience, 1979.~393 p.},
     year = 1979,
   adsurl = {http://adsabs.harvard.edu/abs/1979rpa..book.....R},
   publisher = {John Wiley \& Sons}
}

@ARTICLE{smith+07,
   author = {{Smith}, N. and {Li}, W. and {Foley}, R.~J. and {Wheeler}, J.~C. and 
	{Pooley}, D. and {Chornock}, R. and {Filippenko}, A.~V. and 
	{Silverman}, J.~M. and {Quimby}, R. and {Bloom}, J.~S. and {Hansen}, C.
	},
    title = "{SN 2006gy: Discovery of the Most Luminous Supernova Ever Recorded, Powered by the Death of an Extremely Massive Star like {$\eta$} Carinae}",
  journal = {\apj},
   eprint = {astro-ph/0612617},
     year = 2007,
    month = sep,
   volume = 666,
    pages = {1116--1128},
      doi = {10.1086/519949},
   adsurl = {http://adsabs.harvard.edu/abs/2007ApJ...666.1116S}
}

@ARTICLE{humphreys+davidson+94,
   author = {{Humphreys}, Roberta M. and {Davidson}, Kris},
    title = "{The Luminous Blue Variables: Astrophysical Geysers}",
  journal = {ARA\&A},
     year = 1994,
   volume = {32},
    pages = {303--341},
      doi = {10.1146/annurev.astro.32.1.303},
   adsurl = {https://ui.adsabs.harvard.edu/abs/1994ARA%26A..32..303H}
}

@ARTICLE{vanLoon+05,
   author = {{van Loon}, J.~Th. and {Cioni}, M.-R.~L. and {Zijlstra}, A.~A. and {Loup}, C.},
    title = "{An empirical formula for the mass-loss rates of dust-enshrouded red supergiants and oxygen-rich asymptotic giant branch stars}",
  journal = {\aap},
     year = 2005,
    month = jul,
   volume = {438},
    pages = {273--289},
      doi = {10.1051/0004-6361:20042555},
archivePrefix = {arXiv},
   eprint = {astro-ph/0504379},
 primaryClass = {astro-ph},
   adsurl = {https://ui.adsabs.harvard.edu/abs/2005A%26A...438..273V}
}

@ARTICLE{josselin+07,
   author = {{Josselin}, E. and {Plez}, B.},
    title = "{Atmospheric dynamics and the mass loss process in red supergiant stars}",
  journal = {\aap},
     year = 2007,
    month = jul,
   volume = {469},
    pages = {671--680},
      doi = {10.1051/0004-6361:20070608},
archivePrefix = {arXiv},
   eprint = {0704.0941},
 primaryClass = {astro-ph.SR},
   adsurl = {https://ui.adsabs.harvard.edu/abs/2007A%26A...469..671J}
}

@ARTICLE{crowther+07,
   author = {{Crowther}, Paul A.},
    title = "{Physical Properties of Wolf-Rayet Stars}",
  journal = {ARA\&A},
     year = 2007,
   volume = {45},
    pages = {177--219},
      doi = {10.1146/annurev.astro.45.051806.110615},
archivePrefix = {arXiv},
   eprint = {astro-ph/0610356},
 primaryClass = {astro-ph},
   adsurl = {https://ui.adsabs.harvard.edu/abs/2007ARA%26A..45..177C}
}

@ARTICLE{Sana+12,
   author = {{Sana}, H. and {de Mink}, S.~E. and {de Koter}, A. and {Langer}, N. and {Evans}, C.~J. and {Gieles}, M. and {Gosset}, E. and {Izzard}, R.~G. and {Le Bouquin}, J.-B. and {Schneider}, F.~R.~N. and {Sim{\'o}n-D{\'i}az}, S. and {Tramper}, E. and {van den Heuvel}, E.~P.~J. and {de Wit}, W.-J.},
    title = "{Binary Interaction Dominates the Evolution of Massive Stars}",
  journal = {Science},
     year = 2012,
    month = jul,
   volume = {337},
    pages = {444--446},
      doi = {10.1126/science.1223344},
archivePrefix = {arXiv},
   eprint = {1207.6397},
 primaryClass = {astro-ph.SR},
   adsurl = {https://ui.adsabs.harvard.edu/abs/2012Sci...337..444S}
}

@ARTICLE{smith08tf,
       author = {{Smith}, Nathan and {Chornock}, Ryan and {Li}, Weidong and {Ganeshalingam}, Mohan and {Silverman}, Jeffrey M. and {Foley}, Ryan J. and {Filippenko}, Alexei V. and {Barth}, Aaron J.},
        title = "{SN 2006tf: Precursor Eruptions and the Optically Thick Regime of Extremely Luminous Type IIn Supernovae}",
      journal = {\apj},
         year = 2008,
        month = oct,
       volume = {686},
       number = {1},
        pages = {467-484},
          doi = {10.1086/591021},
archivePrefix = {arXiv},
       eprint = {0804.0042},
 primaryClass = {astro-ph},
       adsurl = {https://ui.adsabs.harvard.edu/abs/2008ApJ...686..467S}
}

@ARTICLE{smith09,
       author = {{Smith}, Nathan and {Silverman}, Jeffrey M. and {Chornock}, Ryan and {Filippenko}, Alexei V. and {Wang}, Xiaofeng and {Li}, Weidong and {Ganeshalingam}, Mohan and {Foley}, Ryan J. and {Rex}, Jacob and {Steele}, Thea N.},
        title = "{Coronal Lines and Dust Formation in SN 2005ip: Not the Brightest, but the Hottest Type IIn Supernova}",
      journal = {\apj},
         year = 2009,
        month = apr,
       volume = {695},
       number = {2},
        pages = {1334-1350},
          doi = {10.1088/0004-637X/695/2/1334},
archivePrefix = {arXiv},
       eprint = {0809.5079},
 primaryClass = {astro-ph},
       adsurl = {https://ui.adsabs.harvard.edu/abs/2009ApJ...695.1334S}
}

@ARTICLE{smith+10,
   author = {{Smith}, N. and {Miller}, A. and {Li}, W. and {Filippenko}, A.~V. and 
  {Silverman}, J.~M. and {Howard}, A.~W. and {Nugent}, P. and 
  {Marcy}, G.~W. and {Bloom}, J.~S. and {Ghez}, A.~M. and {Lu}, J. and 
  {Yelda}, S. and {Bernstein}, R.~A. and {Colucci}, J.~E.},
    title = "{Discovery of Precursor Luminous Blue Variable Outbursts in Two Recent Optical Transients: The Fitfully Variable Missing Links UGC 2773-OT and SN 2009ip}",
  journal = {\aj},
archivePrefix = {arXiv},
   eprint = {0909.4792},
 primaryClass = {astro-ph.SR},
     year = 2010,
    month = apr,
   volume = 139,
    pages = {1451--1467},
      doi = {10.1088/0004-6256/139/4/1451},
   adsurl = {http://adsabs.harvard.edu/abs/2010AJ....139.1451S}
}

@ARTICLE{sa14,
       author = {{Smith}, Nathan and {Arnett}, W. David},
        title = "{Preparing for an Explosion: Hydrodynamic Instabilities and Turbulence in Presupernovae}",
      journal = {\apj},
         year = 2014,
        month = apr,
       volume = {785},
       number = {2},
          eid = {82},
        pages = {82},
          doi = {10.1088/0004-637X/785/2/82},
archivePrefix = {arXiv},
       eprint = {1307.5035},
 primaryClass = {astro-ph.SR},
       adsurl = {https://ui.adsabs.harvard.edu/abs/2014ApJ...785...82S}
}

@ARTICLE{smith+15,
   author = {{Smith}, N. and {Mauerhan}, J.~C. and {Cenko}, S.~B. and {Kasliwal}, M.~M. and 
	{Silverman}, J.~M. and {Filippenko}, A.~V. and {Gal-Yam}, A. and 
	{Clubb}, K.~I. and {Graham}, M.~L. and {Leonard}, D.~C. and 
	{Horst}, J.~C. and {Williams}, G.~G. and {Andrews}, J.~E. and 
	{Kulkarni}, S.~R. and {Nugent}, P. and {Sullivan}, M. and {Maguire}, K. and 
	{Xu}, D. and {Ben-Ami}, S.},
    title = "{PTF11iqb: cool supergiant mass-loss that bridges the gap between Type IIn and normal supernovae}",
  journal = {\mnras},
archivePrefix = {arXiv},
   eprint = {1501.02820},
 primaryClass = {astro-ph.HE},
     year = 2015,
    month = may,
   volume = 449,
    pages = {1876--1896},
      doi = {10.1093/mnras/stv354},
   adsurl = {http://adsabs.harvard.edu/abs/2015MNRAS.449.1876S}
}

@INBOOK{smith+16,
       author = {{Smith}, Nathan},
        title = "{Interacting Supernovae: Types IIn and Ibn}",
    booktitle = {Handbook of Supernovae},
         year = 2017,
    publisher = {Cham, Switzerland: Springer},
        pages = {403},
          doi = {10.1007/978-3-319-21846-5_38},
       adsurl = {https://ui.adsabs.harvard.edu/abs/2017hsn..book..403S}
}

@ARTICLE{smith+17,
   author = {{Smith}, N. and {Kilpatrick}, C.~D. and {Mauerhan}, J.~C. and 
  {Andrews}, J.~E. and {Margutti}, R. and {Fong}, W.-F. and {Graham}, M.~L. and 
  {Zheng}, W. and {Kelly}, P.~L. and {Filippenko}, A.~V. and {Fox}, O.~D.
  },
    title = "{Endurance of SN 2005ip after a decade: X-rays, radio and H{$\alpha$} like SN 1988Z require long-lived pre-supernova mass-loss}",
  journal = {\mnras},
archivePrefix = {arXiv},
   eprint = {1612.02011},
 primaryClass = {astro-ph.HE},
     year = 2017,
    month = apr,
   volume = 466,
    pages = {3021--3034},
      doi = {10.1093/mnras/stw3204},
   adsurl = {http://adsabs.harvard.edu/abs/2017MNRAS.466.3021S}
}

@ARTICLE{Soderberg+06,
       author = {{Soderberg}, A.~M. and {Chevalier}, R.~A. and {Kulkarni}, S.~R. and {Frail}, D.~A.},
        title = "{The Radio and X-Ray Luminous SN 2003bg and the Circumstellar Density Variations around Radio Supernovae}",
      journal = {\apj},
         year = 2006,
        month = nov,
       volume = {651},
       number = {2},
        pages = {1005--1018},
          doi = {10.1086/507571},
archivePrefix = {arXiv},
       eprint = {astro-ph/0512413},
 primaryClass = {astro-ph},
       adsurl = {https://ui.adsabs.harvard.edu/abs/2006ApJ...651.1005S}
}

@ARTICLE{Sollerman+24,
       author = {{Sollerman}, J. and {Yang}, S. and {Schulze}, S. and {Strotjohann}, N.~L. and {Jerkstrand}, A. and {Van Dyk}, S.~D. and {Kool}, E.~C. and {Fransson}, C. and {Gal-Yam}, A. and {Perley}, D.~A.},
        title = "{The dense and non-homogeneous circumstellar medium revealed in radio wavelengths around the Type Ib SN 2019oys}",
      journal = {\aap},
         year = 2024,
        month = jun,
       volume = {686},
          eid = {A157},
        pages = {A157},
          doi = {10.1051/0004-6361/202348761},
archivePrefix = {arXiv},
       eprint = {2311.14409},
 primaryClass = {astro-ph.HE},
       adsurl = {https://ui.adsabs.harvard.edu/abs/2024A&A...686A.157S}
}

@ARTICLE{sramek+80,
   author = {{Sramek}, D. and {van der Hulst}, J.~M. and {Weiler}, K.~W.},
    title = "{Supernova in NGC 6946}",
  journal = {\iaucirc},
     year = 1980,
    month = dec,
   volume = 3557,
   adsurl = {http://adsabs.harvard.edu/abs/1980IAUC.3557....2S}
}

@ARTICLE{taddia+13,
   author = {{Taddia}, F. and {Stritzinger}, M.~D. and {Sollerman}, J. and 
	{Phillips}, M.~M. and {Anderson}, J.~P. and {Boldt}, L. and 
	{Campillay}, A. and {Castell{\'o}n}, S. and {Contreras}, C. and 
	{Folatelli}, G. and {Hamuy}, M. and {Heinrich-Josties}, E. and 
	{Krzeminski}, W. and {Morrell}, N. and {Burns}, C.~R. and {Freedman}, W.~L. and 
	{Madore}, B.~F. and {Persson}, S.~E. and {Suntzeff}, N.~B.},
    title = "{Carnegie Supernova Project: Observations of Type IIn supernovae}",
  journal = {\aap},
archivePrefix = {arXiv},
   eprint = {1304.3038},
     year = 2013,
    month = jul,
   volume = 555,
      eid = {A10},
    pages = {A10},
      doi = {10.1051/0004-6361/201321180},
   adsurl = {http://adsabs.harvard.edu/abs/2013A%26A...555A..10T}
}

@ARTICLE{Tartaglia+25,
       author = {{Tartaglia}, L. and {Elias-Rosa}, N. and {Sand}, D.~J. and {Valenti}, S. and {Andrews}, J.~E. and {Ashall}, C. and {Bostroem}, K.~A. and {Coulter}, D.~A. and {Davis}, K.~W. and {Dickinson}, H.},
        title = "{A Multiwavelength Autopsy of the Interacting Type IIn Supernova 2020ywx: Tracing Its Progenitor Mass-loss History for 100 Yr Before Death}",
      journal = {\apj},
         year = 2025,
        month = mar,
       volume = {983},
       number = {1},
          eid = {62},
        pages = {62},
          doi = {10.3847/1538-4357/adc00a},
archivePrefix = {arXiv},
       eprint = {2409.09560},
 primaryClass = {astro-ph.HE},
       adsurl = {https://ui.adsabs.harvard.edu/abs/2025ApJ...983...62T}
}

@ARTICLE{terreran+16,
   author = {{Terreran}, G. and {Jerkstrand}, A. and {Benetti}, S. and {Smartt}, S.~J. and 
  {Ochner}, P. and {Tomasella}, L. and {Howell}, D.~A. and {Morales-Garoffolo}, A. and 
  {Harutyunyan}, A. and {Kankare}, E. and {Arcavi}, I. and {Cappellaro}, E. and 
  {Elias-Rosa}, N. and {Hosseinzadeh}, G. and {Kangas}, T. and 
  {Pastorello}, A. and {Tartaglia}, L. and {Turatto}, M. and {Valenti}, S. and 
  {Wiggins}, P. and {Yuan}, F.},
    title = "{The multifaceted Type II-L supernova 2014G from pre-maximum to nebular phase}",
  journal = {\mnras},
archivePrefix = {arXiv},
   eprint = {1605.06116},
 primaryClass = {astro-ph.SR},
     year = 2016,
    month = oct,
   volume = 462,
    pages = {137--157},
      doi = {10.1093/mnras/stw1591},
   adsurl = {http://adsabs.harvard.edu/abs/2016MNRAS.462..137T}
}

@ARTICLE{tully+09,
   author = {{Tully}, R.~B. and {Rizzi}, L. and {Shaya}, E.~J. and {Courtois}, H.~M. and 
  {Makarov}, D.~I. and {Jacobs}, B.~A.},
    title = "{The Extragalactic Distance Database}",
  journal = {\aj},
     year = 2009,
    month = aug,
   volume = 138,
    pages = {323--331},
      doi = {10.1088/0004-6256/138/2/323},
   adsurl = {http://adsabs.harvard.edu/abs/2009AJ....138..323T}
}

@ARTICLE{van-dyk+93,
   author = {{van Dyk}, S.~D. and {Weiler}, K.~W. and {Sramek}, R.~A. and 
	{Panagia}, N.},
    title = "{SN 1988Z: The Most Distant Radio Supernova}",
  journal = {\apjl},
     year = 1993,
    month = dec,
   volume = 419,
    pages = {L69},
      doi = {10.1086/187139},
   adsurl = {http://adsabs.harvard.edu/abs/1993ApJ...419L..69V}
}

@ARTICLE{van-dyk+96,
   author = {{van Dyk}, S.~D. and {Weiler}, K.~W. and {Sramek}, R.~A. and 
	{Schlegel}, E.~M. and {Filippenko}, A.~V. and {Panagia}, N. and 
	{Leibundgut}, B.},
    title = "{Type ''IIn'' Supernovae: A Search for Radio Emission}",
  journal = {\aj},
     year = 1996,
    month = mar,
   volume = 111,
    pages = {1271},
      doi = {10.1086/117872},
   adsurl = {http://adsabs.harvard.edu/abs/1996AJ....111.1271V}
}

@ARTICLE{Bietenholz21,
       author = {{Bietenholz}, M.~F. and {Bartel}, N. and {Argo}, M. and
	{Dua}, R. and {Ryder}, S. and {Soderberg}, A.~M.},
        title = "{The Radio Luminosity-risetime Function of Core-collapse Supernovae}",
      journal = {\apj},
         year = 2021,
        month = feb,
       volume = {908},
       number = {1},
          eid = {75},
        pages = {75},
          doi = {10.3847/1538-4357/abccd9},
archivePrefix = {arXiv},
       eprint = {2011.11737},
 primaryClass = {astro-ph.HE},
       adsurl = {https://ui.adsabs.harvard.edu/abs/2021ApJ...908...75B}
}

@ARTICLE{van-dyk+94,
   author = {{van Dyk}, S.~D. and {Weiler}, K.~W. and {Sramek}, R.~A. and 
  {Rupen}, M.~P. and {Panagia}, N.},
    title = "{SN 1993J: The early radio emission and evidence for a changing presupernova mass-loss rate}",
  journal = {\apjl},
     year = 1994,
    month = sep,
   volume = 432,
    pages = {L115--L118},
      doi = {10.1086/187525},
   adsurl = {http://adsabs.harvard.edu/abs/1994ApJ...432L.115V}
}

@INPROCEEDINGS{weiler+01,
   author = {{Weiler}, K.~W. and {Panagia}, N. and {Sramek}, R.~A. and {van Dyk}, S.~D. and 
	{Montes}, M.~J. and {Lacey}, C.~K.},
    title = "{Radio supernovae and GRB 980425}",
booktitle = {Supernovae and Gamma-Ray Bursts: the Greatest Explosions since the Big Bang},
   series = {Supernovae and Gamma-Ray Bursts: the Greatest Explosions since the Big Bang},
     year = 2001,
   volume = 13,
   eprint = {astro-ph/0002501},
   editor = {{Livio}, M. and {Panagia}, N. and {Sahu}, K.},
    pages = {198--217},
   adsurl = {http://adsabs.harvard.edu/abs/2001sgrb.conf..198W}
}

@ARTICLE{weiler+02,
   author = {{Weiler}, K.~W. and {Panagia}, N. and {Montes}, M.~J. and {Sramek}, R.~A.
	},
    title = "{Radio Emission from Supernovae and Gamma-Ray Bursters}",
  journal = {\araa},
     year = 2002,
   volume = 40,
    pages = {387--438},
      doi = {10.1146/annurev.astro.40.060401.093744},
   adsurl = {http://adsabs.harvard.edu/abs/2002ARA%26A..40..387W}
}

@INPROCEEDINGS{weiler+82,
   author = {{Weiler}, K.~W. and {Sramek}, R.~A. and {van der Hulst}, J.~M. and 
	{Panagia}, N.},
    title = "{Radio supernovae}",
booktitle = {NATO Advanced Science Institutes (ASI) Series C},
     year = 1982,
   series = {NATO Advanced Science Institutes (ASI) Series C},
   volume = 90,
   editor = {{Rees}, M.~J. and {Stoneham}, R.~J.},
    month = nov,
    pages = {281--291},
   adsurl = {http://adsabs.harvard.edu/abs/1982ASIC...90..281W}
}

@ARTICLE{weiler+86,
   author = {{Weiler}, K.~W. and {Sramek}, R.~A. and {Panagia}, N. and {van der Hulst}, J.~M. and 
	{Salvati}, M.},
    title = "{Radio supernovae}",
  journal = {\apj},
     year = 1986,
    month = feb,
   volume = 301,
    pages = {790--812},
      doi = {10.1086/163944},
   adsurl = {http://adsabs.harvard.edu/abs/1986ApJ...301..790W}
}

@ARTICLE{weiler+90,
   author = {{Weiler}, K.~W. and {Panagia}, N. and {Sramek}, R.~A.},
    title = "{Radio emission from supernovae. II - SN 1986J: A different kind of type II}",
  journal = {\apj},
     year = 1990,
    month = dec,
   volume = 364,
    pages = {611--625},
      doi = {10.1086/169444},
   adsurl = {http://adsabs.harvard.edu/abs/1990ApJ...364..611W}
}

@ARTICLE{weiler+91,
   author = {{Weiler}, K.~W. and {van Dyk}, S.~D. and {Discenna}, J.~L. and 
	{Panagia}, N. and {Sramek}, R.~A.},
    title = "{The 10 year radio light curves for SN 1979C}",
  journal = {\apj},
     year = 1991,
    month = oct,
   volume = 380,
    pages = {161--166},
      doi = {10.1086/170571},
   adsurl = {http://adsabs.harvard.edu/abs/1991ApJ...380..161W}
}

@ARTICLE{weiler+92,
   author = {{Weiler}, K.~W. and {van Dyk}, S.~D. and {Panagia}, N. and {Sramek}, R.~A.
	},
    title = "{Full evolution of the 6 and 20 centimeter radio emission from SN 1980K}",
  journal = {\apj},
     year = 1992,
    month = oct,
   volume = 398,
    pages = {248--253},
      doi = {10.1086/171852},
   adsurl = {http://adsabs.harvard.edu/abs/1992ApJ...398..248W}
}

@ARTICLE{williams+02,
   author = {{Williams}, C.~L. and {Panagia}, N. and {Van Dyk}, S.~D. and 
	{Lacey}, C.~K. and {Weiler}, K.~W. and {Sramek}, R.~A.},
    title = "{Radio Emission from SN 1988Z and Very Massive Star Evolution}",
  journal = {\apj},
   eprint = {astro-ph/0208190},
     year = 2002,
    month = dec,
   volume = 581,
    pages = {396--403},
      doi = {10.1086/344087},
   adsurl = {http://adsabs.harvard.edu/abs/2002ApJ...581..396W}
}

@ARTICLE{Perley+11,
   author = {{Perley}, R.~A. and {Chandler}, C.~J. and {Butler}, B.~J. and {Wrobel}, J.~M.},
    title = "{The Expanded Very Large Array: A New Telescope for Centimeter- to Meter-Wavelength Astronomy}",
  journal = {\apjs},
     year = 2011,
    month = jul,
   volume = {194},
    number = {1},
     pages = {25},
      doi = {10.1088/0067-0049/194/1/25},
   adsurl = {https://ui.adsabs.harvard.edu/abs/2011ApJS..194...25P}
}

@ARTICLE{Dewdney+09,
   author = {{Dewdney}, P.~E. and {Hall}, P.~J. and {Schilizzi}, R.~T. and {Lazio}, T.~J.~W.~W.},
    title = "{The Square Kilometre Array}",
  journal = {Proceedings of the IEEE},
     year = 2009,
    month = aug,
   volume = {97},
    number = {8},
     pages = {1482--1496},
      doi = {10.1109/JPROC.2009.2021005},
   adsurl = {https://ui.adsabs.harvard.edu/abs/2009IEEEP..97.1482D}
}

@ARTICLE{chandra+18,
       author = {{Chandra}, Poonam},
        title = "{Circumstellar Interaction in Supernovae in Dense Environments---An Observational Perspective}",
      journal = {Space Science Reviews},
         year = 2018,
        month = feb,
       volume = {214},
       number = {2},
          eid = {27},
        pages = {27},
          doi = {10.1007/s11214-017-0461-6},
       adsurl = {https://ui.adsabs.harvard.edu/abs/2018SSRv..214...27C}
}

@ARTICLE{fraser+20,
       author = {{Fraser}, Morgan},
        title = "{Supernovae and transients with circumstellar interaction}",
      journal = {Royal Society Open Science},
         year = 2020,
        month = jul,
       volume = {7},
       number = {7},
          eid = {200467},
        pages = {200467},
          doi = {10.1098/rsos.200467},
       adsurl = {https://ui.adsabs.harvard.edu/abs/2020RSOS....7.200467F}
}

@ARTICLE{boisseau+21,
       author = {{Boisseau}, Thierry and {Irwin}, Christopher M.},
        title = "{The Radio Luminosity-Risetime Function of Core-collapse Supernovae}",
      journal = {\apj},
         year = 2021,
        month = feb,
       volume = {908},
       number = {2},
          eid = {75},
        pages = {75},
          doi = {10.3847/1538-4357/abd5b4},
archivePrefix = {arXiv},
       eprint = {2106.09737},
 primaryClass = {astro-ph.HE},
       adsurl = {https://ui.adsabs.harvard.edu/abs/2021ApJ...908...75B}
}

@ARTICLE{Tartaglia+21,
       author = {{Tartaglia}, Leonardo and {Sarin}, Nikhil and {Smith}, Nathan and {Bostroem}, K.~A. and {Kochanek}, C.~S. and {Shappee}, B.~J. and {Holoien}, T.~W. -S. and {Dong}, Subo and {Stritzinger}, M.~D. and {Brown}, Jonathan S. and {Prieto}, Jos{\'e}~L. and {Shields}, Joseph and {Thompson}, Todd A.},
        title = "{Multiwavelength View of the Type IIn Zoo: Optical to X-Ray Emission Model of Interaction-powered Supernovae}",
      journal = {\apj},
         year = 2021,
        month = jun,
       volume = {914},
       number = {1},
          eid = {64},
        pages = {64},
          doi = {10.3847/1538-4357/abf7e2},
archivePrefix = {arXiv},
       eprint = {2104.06389},
 primaryClass = {astro-ph.HE},
       adsurl = {https://ui.adsabs.harvard.edu/abs/2021ApJ...914...64T}
}

@ARTICLE{taddia+19,
       author = {{Taddia}, F. and {Sollerman}, J. and {Leloudas}, G. and {Ergon}, M. and {Nyholm}, A. and {Fremling}, C. and {Karamehmetoglu}, E. and {Roy}, R. and {Fransson}, C. and {Barbarino}, C. and {Dessart}, L. and {Moriya}, T.~J. and {Smartt}, S.~J. and {Tomasella}, L.},
        title = "{Type IIn Supernova Light Curves Powered by Forward and Reverse Shocks}",
      journal = {\apj},
         year = 2019,
        month = aug,
       volume = {884},
       number = {2},
          eid = {87},
        pages = {87},
          doi = {10.3847/1538-4357/ab2d3e},
archivePrefix = {arXiv},
       eprint = {1906.05812},
 primaryClass = {astro-ph.HE},
       adsurl = {https://ui.adsabs.harvard.edu/abs/2019ApJ...884...87T}
}

@ARTICLE{murase+19,
       author = {{Murase}, Kohta and {Thompson}, Todd A. and {Lacki}, Brian C. and {Beacom}, John F.},
        title = "{High-energy Emission from Interacting Supernovae: New Constraints on Cosmic-Ray Acceleration in Dense Circumstellar Environments}",
      journal = {\apj},
         year = 2019,
        month = mar,
       volume = {874},
       number = {2},
          eid = {80},
        pages = {80},
          doi = {10.3847/1538-4357/ab0e6e},
archivePrefix = {arXiv},
       eprint = {1807.01460},
 primaryClass = {astro-ph.HE},
       adsurl = {https://ui.adsabs.harvard.edu/abs/2019ApJ...874...80M}
}

@ARTICLE{chevalier-fransson+06araa,
       author = {{Chevalier}, Roger A. and {Fransson}, Claes},
        title = "{The Interaction of Supernovae with the Circumstellar Medium}",
      journal = {ARA\&A},
         year = 2006,
        month = sep,
       volume = {44},
        pages = {17--48},
          doi = {10.1146/annurev.astro.44.051805.110103},
       adsurl = {https://ui.adsabs.harvard.edu/abs/2006ARA%26A..44...17C}
}

@ARTICLE{dessart+23,
       author = {{Dessart}, L. and {Jacobson-Gal{\'a}n}, W.~V.},
        title = "{Using spectral modeling to break light-curve degeneracies of type II supernovae interacting with circumstellar material}",
      journal = {\aap},
         year = 2023,
        month = sep,
       volume = {677},
          eid = {A105},
        pages = {A105},
          doi = {10.1051/0004-6361/202346754},
       adsurl = {https://ui.adsabs.harvard.edu/abs/2023A&A...677A.105D}
}

@ARTICLE{morozova+17,
       author = {{Morozova}, V.~I. and {Piro}, A.~L. and {Valenti}, S.},
        title = "{Unifying Type II Supernova Light Curves with Dense Circumstellar Material}",
      journal = {\apj},
         year = 2017,
        month = mar,
       volume = {838},
       number = {1},
          eid = {28},
        pages = {28},
          doi = {10.3847/1538-4357/aa6251},
archivePrefix = {arXiv},
       eprint = {1610.08054},
 primaryClass = {astro-ph.HE},
       adsurl = {https://ui.adsabs.harvard.edu/abs/2017ApJ...838...28M}
}

@INPROCEEDINGS{Selina+18ngVLA,
       author = {{Selina}, Robert and {Murphy}, Eric J. and {McKinnon}, Mark M. and {Beasley}, Anthony and {Butler}, Bryan and {Carilli}, Chris and {Clark}, Barry and {Durand}, Steven and {Erickson}, Alan and {Hiriart}, Rafael and {Grammer}, Wes and {Jackson}, James and {Kent}, Brian and {Mason}, Brian and {Morgan}, Matthew and {Yeste Ojeda}, Omar and {Rosero}, Viviana and {Shillue}, William and {Sturgis}, Silver and {Urbain}, Denis},
        title = "{The ngVLA Reference Design}",
    booktitle = {Science with a Next Generation Very Large Array},
       editor = {{Murphy}, Eric J.},
      series = {ASP Conf. Ser.},
       volume = {517},
        pages = {15},
    publisher = {Astronomical Society of the Pacific},
      address = {San Francisco, CA},
         year = 2018,
        month = dec,
archivePrefix = {arXiv},
       eprint = {1810.08197},
 primaryClass = {astro-ph.IM},
       adsurl = {https://ui.adsabs.harvard.edu/abs/2018ASPC..517...15S}
}

@ARTICLE{Nakano6505,
   author = {{Nakano}, S. and {Kushida}, R. and {Koshida}, Y. and {Garnavich}, P. and
	{Challis}, P. and {Kirshner}, R. and {Hergenrother}, C. and
	{Brown}, W. and {Marzke}, R. and {Odewahn}, S.},
    title = "{Supernova 1996bu in NGC 3631}",
  journal = {\iaucirc},
     year = 1996,
    month = nov,
   volume = 6505,
   adsurl = {http://adsabs.harvard.edu/abs/1996IAUC.6505....1N}
}

@ARTICLE{Li6829,
   author = {{Li}, W.-D. and {Li}, C. and {Filippenko}, A.~V. and {Moran}, E.~C.},
    title = "{Supernova 1998S in NGC 3877}",
  journal = {\iaucirc},
     year = 1998,
    month = mar,
   volume = 6829,
   adsurl = {http://adsabs.harvard.edu/abs/1998IAUC.6829....1L}
}

@ARTICLE{Filippenko6830,
   author = {{Filippenko}, A.~V. and {Moran}, E.~C.},
    title = "{Supernova 1998S in NGC 3877}",
  journal = {\iaucirc},
     year = 1998,
    month = mar,
   volume = 6830,
   adsurl = {http://adsabs.harvard.edu/abs/1998IAUC.6830....2F}
}

@ARTICLE{Jha7379,
   author = {{Jha}, S. and {Challis}, P. and {Kirshner}, R. and {Calkins}, M.},
    title = "{Supernova 2000P in NGC 4965}",
  journal = {\iaucirc},
     year = 2000,
    month = mar,
   volume = 7379,
   adsurl = {http://adsabs.harvard.edu/abs/2000IAUC.7379....1J}
}

@ARTICLE{Jha7381,
   author = {{Jha}, S. and {Challis}, P. and {Kirshner}, R.},
    title = "{Supernova 2000P in NGC 4965}",
  journal = {\iaucirc},
     year = 2000,
    month = mar,
   volume = 7381,
   adsurl = {http://adsabs.harvard.edu/abs/2000IAUC.7381....2J}
}

@ARTICLE{Yu7476,
   author = {{Yu}, C. and {Li}, W.~D.},
    title = "{Supernova 2000dc in ESO 527-G019}",
  journal = {\iaucirc},
     year = 2000,
    month = aug,
   volume = 7476,
   adsurl = {http://adsabs.harvard.edu/abs/2000IAUC.7476....1Y}
}

@ARTICLE{Leonard7483,
   author = {{Leonard}, D.~C. and {Li}, W.~D. and {Filippenko}, A.~V.},
    title = "{Supernovae 2000dc, 2000dd, and 2000df}",
  journal = {\iaucirc},
     year = 2000,
    month = aug,
   volume = 7483,
   adsurl = {http://adsabs.harvard.edu/abs/2000IAUC.7483....3L}
}

@ARTICLE{Modjaz7682,
   author = {{Modjaz}, M. and {Li}, W.~D.},
    title = "{Supernova 2001do in UGC 11459}",
  journal = {\iaucirc},
     year = 2001,
    month = aug,
   volume = 7682,
   adsurl = {http://adsabs.harvard.edu/abs/2001IAUC.7682....1M}
}

@ARTICLE{Chornock7699,
   author = {{Chornock}, R. and {Modjaz}, M. and {Filippenko}, A.~V.},
    title = "{Supernova 2001do in UGC 11459}",
  journal = {\iaucirc},
     year = 2001,
    month = aug,
   volume = 7699,
   adsurl = {http://adsabs.harvard.edu/abs/2001IAUC.7699....1C}
}

@ARTICLE{Boles275,
   author = {{Boles}, T. and {Nakano}, S. and {Itagaki}, K.},
    title = "{Supernova 2005ip in NGC 2906}",
  journal = {Central Bureau Electronic Telegrams},
     year = 2005,
    month = nov,
   volume = 275,
   adsurl = {http://adsabs.harvard.edu/abs/2005CBET..275....1B}
}

@ARTICLE{Modjaz276,
   author = {{Modjaz}, M. and {Kirshner}, R. and {Challis}, P.},
    title = "{Supernova 2005ip in NGC 2906}",
  journal = {Central Bureau Electronic Telegrams},
     year = 2005,
    month = nov,
   volume = 276,
   adsurl = {http://adsabs.harvard.edu/abs/2005CBET..276....1M}
}

@ARTICLE{Smith695,
       author = {{Smith}, Nathan and {Silverman}, Jeffrey M. and {Chornock}, Ryan and {Filippenko}, Alexei V. and {Wang}, Xiaofeng and {Li}, Weidong and {Ganeshalingam}, Mohan and {Foley}, Ryan J. and {Rex}, Jacob and {Steele}, Thea N.},
        title = "{Coronal Lines and Dust Formation in SN 2005ip: Not the Brightest, but the Hottest Type IIn Supernova}",
      journal = {\apj},
         year = 2009,
        month = apr,
       volume = {695},
       number = {2},
        pages = {1334--1350},
          doi = {10.1088/0004-637X/695/2/1334},
       adsurl = {https://ui.adsabs.harvard.edu/abs/2009ApJ...695.1334S}
}

@ARTICLE{Lee412,
   author = {{Lee}, E. and {Li}, W.},
    title = "{Supernova 2006am in NGC 5630}",
  journal = {Central Bureau Electronic Telegrams},
     year = 2006,
    month = feb,
   volume = 412,
   adsurl = {http://adsabs.harvard.edu/abs/2006CBET..412....1L}
}

@ARTICLE{Blondin8680,
   author = {{Blondin}, S. and {Modjaz}, M. and {Kirshner}, R. and {Challis}, P.},
    title = "{Supernovae 2006R--2006T, 2006V, 2006X, 2006al, 2006am, 2006an}",
  journal = {\iaucirc},
     year = 2006,
    month = feb,
   volume = 8680,
   adsurl = {http://adsabs.harvard.edu/abs/2006IAUC.8680....1B}
}

@ARTICLE{Thrasher1507,
   author = {{Thrasher}, P. and {Li}, W. and {Filippenko}, A.~V.},
    title = "{Supernova 2008fq in NGC 6907}",
  journal = {Central Bureau Electronic Telegrams},
     year = 2008,
    month = sep,
   volume = 1507,
   adsurl = {http://adsabs.harvard.edu/abs/2008CBET.1507....1T}
}

@ARTICLE{Quinn1510,
   author = {{Quinn}, J. and {Baade}, D. and {Clocchiatti}, A. and {Maund}, J. and
	{Patat}, F. and {Wang}, L.},
    title = "{Supernova 2008fq}",
  journal = {Central Bureau Electronic Telegrams},
     year = 2008,
    month = sep,
   volume = 1510,
   adsurl = {http://adsabs.harvard.edu/abs/2008CBET.1510....1Q}
}

@ARTICLE{Mauerhan430,
   author = {{Mauerhan}, J.~C. and
	{Smith}, N. and
	{Filippenko}, A.~V. and
	{Blanchard}, K.~B. and
	{Blanchard}, P.~K. and
	{Casper}, C.~F.~E. and
	{Cenko}, S.~B. and
	{Clubb}, K.~I. and
	{Cohen}, D.~P. and
	{Fuller}, K.~L. and
	{Li}, G.~Z. and
	{Silverman}, J.~M.},
    title = "{The Unprecedented 2012 Outburst of SN 2009ip: A Luminous Blue Variable Star Becomes a True Supernova}",
  journal = {\mnras},
     year = 2013,
    month = mar,
   volume = {430},
    pages = {1801--1810},
      doi = {10.1093/mnras/stt009},
   adsurl = {https://ui.adsabs.harvard.edu/abs/2013MNRAS.430.1801M}
}

@ARTICLE{Pastorello767,
   author = {{Pastorello}, A. and {Cappellaro}, E. and {Inserra}, C. and
	{Smartt}, S.~J. and {Pignata}, G. and {Benetti}, S. and {Valenti}, S. and
	{Fraser}, M. and {Tak{\'a}ts}, K. and {Benitez}, S. and {Botticella}, M.~T. and
	{Brimacombe}, J. and {Bufano}, F. and {Cellier-Holzem}, F. and
	{Costado}, M.~T. and {Cupani}, G. and {Curtis}, I. and {Elias-Rosa}, N. and
	{Ergon}, M. and {Fynbo}, J.~P.~U. and {Hambsch}, F.-J. and {Hamuy}, M. and
	{Harutyunyan}, A. and {Ivarson}, K.~M. and {Kankare}, E. and
	{Martin}, J.~C. and {Kotak}, R. and {LaCluyze}, A.~P. and {Maguire}, K. and
	{Mattila}, S. and {Maza}, J. and {McCrum}, M. and {Miluzio}, M. and
	{Norgaard-Nielsen}, H.~U. and {Nysewander}, M.~C. and {Ochner}, P. and
	{Pan}, Y.-C. and {Pumo}, M.~L. and {Reichart}, D.~E. and {Tan}, T.~G. and
	{Taubenberger}, S. and {Tomasella}, L. and {Turatto}, M. and
	{Wright}, D.},
    title = "{Interacting Supernovae and Supernova Impostors: SN 2009ip, is this the End?}",
  journal = {\apj},
     year = 2013,
    month = apr,
   volume = 767,
      eid = {1},
    pages = {1},
      doi = {10.1088/0004-637X/767/1/1},
   adsurl = {http://adsabs.harvard.edu/abs/2013ApJ...767....1P}
}

@ARTICLE{Fraser433,
   author = {{Fraser}, M. and {Inserra}, C. and {Jerkstrand}, A. and {Kotak}, R. and
	{Pignata}, G. and {Benetti}, S. and {Botticella}, M. and {Bufano}, F. and
	{Childress}, M. and {Mattila}, S. and {Pastorello}, A. and {Smartt}, S.~J. and
	{Turatto}, M. and {Yuan}, F. and {Anderson}, J.~P. and {Bayliss}, D.~D.~R. and
	{Bauer}, F.~E. and {Chen}, T. and {Forster Buron}, F. and {Gal-Yam}, A. and
	{Haislip}, J.~B. and {Knapic}, C. and {Le Guillou}, L. and {Marchi}, S. and
	{Mazzali}, P. and {Molinaro}, M. and {Moore}, J.~P. and {Reichart}, D. and
	{Smareglia}, R. and {Smith}, K.~W. and {Sternberg}, A. and {Sullivan}, M. and
	{Tak{\'a}ts}, K. and {Tucker}, B.~E. and {Valenti}, S. and {Yaron}, O. and
	{Young}, D.~R. and {Zhou}, G.},
    title = "{SN 2009ip a la PESSTO: No Evidence for Core Collapse Yet}",
  journal = {\mnras},
     year = 2013,
    month = aug,
   volume = {433},
    pages = {1312--1337},
      doi = {10.1093/mnras/stt813},
   adsurl = {https://ui.adsabs.harvard.edu/abs/2013MNRAS.433.1312F}
}

@ARTICLE{Monard1867,
   author = {{Monard}, L.~A.~G.},
    title = "{Supernova 2009hd in M66}",
  journal = {Central Bureau Electronic Telegrams},
     year = 2009,
    month = jul,
   volume = 1867,
   adsurl = {http://adsabs.harvard.edu/abs/2009CBET.1867....1M}
}

@ARTICLE{EliasRosa742,
   author = {{Elias-Rosa}, N. and {Van Dyk}, S.~D. and {Li}, W. and {Silverman}, J.~M. and
	{Foley}, R.~J. and {Ganeshalingam}, M. and {Mauerhan}, J.~C. and
	{Kankare}, E. and {Jha}, S. and {Filippenko}, A.~V. and {Beckman}, J.~E. and
	{Berger}, E. and {Cuillandre}, J.-C. and {Smith}, N.},
    title = "{The Massive Progenitor of the Possible Type II-Linear Supernova 2009hd in Messier 66}",
  journal = {\apj},
     year = 2011,
    month = nov,
   volume = 742,
      eid = {6},
    pages = {6},
      doi = {10.1088/0004-637X/742/1/6},
   adsurl = {http://adsabs.harvard.edu/abs/2011ApJ...742....6E}
}

@ARTICLE{Nakano2006,
   author = {{Nakano}, S. and {Yusa}, T. and {Kadota}, K.},
    title = "{Supernova 2009kr in NGC 1832}",
  journal = {Central Bureau Electronic Telegrams},
     year = 2009,
    month = nov,
   volume = 2006,
   adsurl = {http://adsabs.harvard.edu/abs/2009CBET.2006....1N}
}

@ARTICLE{Li2042,
   author = {{Li}, W. and {Filippenko}, A.~V. and {Miller}, A.~A. and {Cuillandre}, J.-C. and
	{Elias-Rosa}, N. and {van Dyk}, S.~D.},
    title = "{Supernova 2009kr in NGC 1832}",
  journal = {Central Bureau Electronic Telegrams},
     year = 2009,
    month = nov,
   volume = 2042,
   adsurl = {http://adsabs.harvard.edu/abs/2009CBET.2042....1L}
}

@ARTICLE{Maza2544,
   author = {{Maza}, J. and {Hamuy}, M. and {Antezana}, R. and {Gonzalez}, L. and
	{Cartier}, R. and {Forster}, F. and {Silva}, S. and {Carrasco}, F. and
	{Pignata}, G. and {Cifuentes}, M. and {Conuel}, B. and {Folatelli}, G. and
	{Reichart}, D. and {Ivarsen}, K. and {Haislip}, J. and {Crain}, A. and
	{Foster}, D. and {Nysewander}, M. and {Lacluyze}, A.},
    title = "{Supernovae 2010jp and 2010jq}",
  journal = {Central Bureau Electronic Telegrams},
     year = 2010,
    month = nov,
   volume = 2544,
   adsurl = {http://adsabs.harvard.edu/abs/2010CBET.2544....1M}
}

@ARTICLE{Challis2548,
   author = {{Challis}, P. and {Kirshner}, R. and {Smith}, N.},
    title = "{Supernovae 2010jp, 2010jr, and 2010js}",
  journal = {Central Bureau Electronic Telegrams},
     year = 2010,
    month = nov,
   volume = 2548,
   adsurl = {http://adsabs.harvard.edu/abs/2010CBET.2548....1C}
}

@ARTICLE{Pignata2623,
   author = {{Pignata}, G. and {Cifuentes}, M. and {Maza}, J. and {Hamuy}, M. and
	{Antezana}, R. and {Gonzalez}, L. and {Cartier}, R. and {Forster}, F. and
	{Silva}, S. and {Carrasco}, F. and {Gonzalez}, P. and {Conuel}, B. and
	{Folatelli}, G. and {Reichart}, D. and {Ivarsen}, K. and {Haislip}, J. and
	{Crain}, A. and {Foster}, D. and {Nysewander}, M. and {Lacluyze}, A. and
	{Stritzinger}, M. and {Prieto}, J.~L. and {Morrell}, N.},
    title = "{Supernova 2011A in NGC 4902}",
  journal = {Central Bureau Electronic Telegrams},
     year = 2011,
    month = jan,
   volume = 2623,
   adsurl = {http://adsabs.harvard.edu/abs/2011CBET.2623....1P}
}

@ARTICLE{Boles2851,
   author = {{Boles}, T. and {Pastorello}, A. and {Stanishev}, V. and {Smartt}, S.~J. and
	{Fraser}, M. and {Lindborg}, M.},
    title = "{Psn J10081059+5150570 in UGC 5460}",
  journal = {Central Bureau Electronic Telegrams},
     year = 2011,
    month = oct,
   volume = 2851,
   adsurl = {http://adsabs.harvard.edu/abs/2011CBET.2851....1B}
}

@ARTICLE{Prieto3749,
   author = {{Prieto}, J.~L. and {Shappee}, B.~J. and {Stanek}, K.~Z. and {Kochanek}, C.~S. and
	{Szczygiel}, D. and {Beacom}, J.~D. and {Pojmanski}, G. and {Rosing}, W. and
	{Hawkins}, E. and {Ross}, R. and {Elphick}, D. and {Mullins}, D. and {Walker}, Z.},
    title = "{ASAS-SN and Swift follow-up of PSN J10081059+5150570: An Unusual Type IIn Supernova ?}",
  journal = {The Astronomer's Telegram},
     year = 2011,
    month = nov,
   volume = 3749,
   adsurl = {http://adsabs.harvard.edu/abs/2011ATel.3749....1P}
}

@ARTICLE{Mauerhan431,
   author = {{Mauerhan}, J.~C. and {Smith}, N. and {Silverman}, J.~M. and
	{Filippenko}, A.~V. and {Morgan}, A.~N. and {Cenko}, S.~B. and
	{Ganeshalingam}, M. and {Bloom}, J.~S. and {Matheson}, T. and {Milne}, P.},
    title = "{SN 2011ht: confirming a class of interacting supernovae with plateau light curves (Type IIn-P)}",
  journal = {\mnras},
     year = 2013,
    month = may,
   volume = 431,
    pages = {2599--2611},
      doi = {10.1093/mnras/stt360},
   adsurl = {http://adsabs.harvard.edu/abs/2013MNRAS.431.2599M}
}

@ARTICLE{Parrent3510,
   author = {{Parrent}, J. and {Levitan}, D. and {Howell}, A. and {Thomas}, R.~C. and
	{Nugent}, P. and {Sullivan}, M. and {Kasliwal}, M. and {Ofek}, E.~O. and
	{Quimby}, R. and {Ben-Ami}, S. and {Xu}, D. and {Arcavi}, I. and
	{Gal-Yam}, A. and {Cenko}, C.~B. and {Li}, W. and {Filippenko}, A.~V.},
    title = "{PTF discovers a young type IIn SN in NGC 151}",
  journal = {The Astronomer's Telegram},
     year = 2011,
    month = jul,
   volume = 3510,
   adsurl = {http://adsabs.harvard.edu/abs/2011ATel.3510....1P}
}

@ARTICLE{Smith449,
       author = {{Smith}, N. and {Mauerhan}, J.~C. and {Cenko}, S.~B. and {Kasliwal}, M.~M. and
	{Silverman}, J.~M. and {Filippenko}, A.~V. and {Gal-Yam}, A. and
	{Clubb}, K.~I. and {Graham}, M.~L. and {Leonard}, D.~C. and
	{Horst}, J.~C. and {Williams}, G.~G. and {Andrews}, J.~E. and
	{Kulkarni}, S.~R. and {Nugent}, P. and {Sullivan}, M. and {Maguire}, K. and
	{Xu}, D. and {Ben-Ami}, S.},
        title = "{PTF11iqb: cool supergiant mass-loss that bridges the gap between Type IIn and normal supernovae}",
      journal = {\mnras},
         year = 2015,
        month = may,
       volume = 449,
        pages = {1876--1896},
          doi = {10.1093/mnras/stv354},
       adsurl = {http://adsabs.harvard.edu/abs/2015MNRAS.449.1876S}
}

@ARTICLE{Jin4051,
   author = {{Jin}, Z.-w. and {Gao}, X. and {Migotto}, K. and {Fremling}, C. and
	{Nyholm}, A. and {Taddia}, F. and {Karamehmetoglu}, E. and {Sollerman}, J. and
	{Elias-Rosa}, N. and {Galbany}, L. and {Inserra Queen's University}, C. and
	{Maguire}, K. and {Smartt}, S.~J. and {Smith}, K.~W. and {Sullivan}, M. and
	{Valenti}, S. and {Yaron}, O. and {Young}, D. and {Manulis}, I.},
    title = "{Supernova 2015D in NGC 5020 = Psn J13124116+1236018}",
  journal = {Central Bureau Electronic Telegrams},
     year = 2015,
    month = jan,
   volume = 4051,
   adsurl = {http://adsabs.harvard.edu/abs/2015CBET.4051....1J}
}

@ARTICLE{Zhang6939,
   author = {{Zhang}, J. and {Wang}, X.},
    title = "{Spectroscopic Classification of PSN J13522411+3941286 as a Type IIn Supernova}",
  journal = {The Astronomer's Telegram},
     year = 2015,
    month = jan,
   volume = 6939,
   adsurl = {http://adsabs.harvard.edu/abs/2015ATel.6939....1Z}
}

\end{document}